\documentclass[12pt]{iopart}
\usepackage{amsbsy}
\usepackage{amssymb}

\usepackage[colorlinks,bookmarks=false,citecolor=blue,linkcolor=red,urlcolor=blue]{hyperref}
\usepackage[all]{xy}
\usepackage{graphicx}
\def\bc{\begin{center}} \def\ec{\end{center}}
\def\be{\begin{equation}} \def\ee{\end{equation}}
\def\bea{\begin{eqnarray}} \def\eea{\end{eqnarray}}

\newcommand{\bra}[1]{\langle\,#1\,|} 
\newcommand{\ket}[1]{|#1\,\rangle}

\newcommand{\eg}{\textit{e.g.} }

\newcommand{\up}{\uparrow} \newcommand{\dw}{\downarrow}

\begin{document}
\title[Participation spectroscopy and entanglement Hamiltonian of quantum spin models]{Participation spectroscopy and entanglement Hamiltonian of quantum spin models}
\author{David J. Luitz, Nicolas Laflorencie, Fabien Alet}
\ead{luitz@irsamc.ups-tlse.fr, laflo@irsamc.ups-tlse.fr,alet@irsamc.ups-tlse.fr}
\address{Laboratoire de Physique Th\'eorique, IRSAMC, Universit\'e de Toulouse,{CNRS, 31062 Toulouse, France}}
\begin{abstract}
Shannon-R\'enyi entropies and associated participation spectra quantify how much a many-body wave-function is localized in a given configuration basis. Using these tools, we present an analysis of the ground-state wave functions of various quantum spin systems in one and two dimensions. General ideas and a review of the current status of this field are first given, with a particular emphasis on universal subleading terms characterizing different quantum phases of matter, and associated transitions. We highlight the connection with the related entanglement entropies and spectra when this is possible.

 In a second part, new results are presented for the participation spectra of interacting spin models, mostly based on quantum Monte Carlo simulations, but also using perturbation theory in some cases. For full antiferromagnetic systems, participation spectra are analyzed in terms of ferromagnetic domain walls which experience a repulsive pairwise interaction. This confinement potential is either linear for long-range N\'eel order, or logarithmic for quasi-long-range order. The case of subsystems is also analyzed in great detail for a 2d dimerized Heisenberg model undergoing a quantum phase transition between a  gapped paramagnet and a N\'eel phase. Participation spectra of line shaped (1d) sub-systems are quantitatively compared with finite temperature participation spectra of ansatz effective boundary (1d) entanglement Hamiltonians. While short-range models describe almost perfectly the gapped side, the N\'eel regime is best compared using long-range effective Hamiltonians. Spectral comparisons performed using Kullback-Leibler divergences, a tool potentially useful for entanglement spectra, provide a quantitative way to identify both the best boundary entanglement Hamiltonian and temperature.
\end{abstract}


\submitto{J. Stat. Mech.}
\section{Introduction}

This work aims at studying the statistical properties of the coefficients of the ground-state
wave-function of several interacting (many-body) quantum spin systems, searching for universal
signatures of quantum phases of matter and quantum phase transitions. The main objects that will be considered are the {\it R\'enyi entropies} 
\be
\label{eq:SR1}
S_q = \frac{1}{1-q} \ln \sum_i p_i^q
\ee
with the particular case of the {\it Shannon entropy} 
\be
\label{eq:SR2} 
S_1 = \lim_{q\rightarrow 1} S_q = -\sum_i p_i \ln p_i. 
\ee
We adopt the generic name of Shannon-R\'enyi (SR) entropies in the following to denote $S_q$, and consider only $q\geq 0$ for the rest of this work.
A more complete characterization is given by what we dub the ``participation spectrum" (PS), a set of positive numbers defined by 
\be
\label{eq:PS}
\epsilon_i = -\ln(p_i).
\ee

In the equations above, $p_i=|a_i|^2$ denotes the probability (or ``participation") of a basis state
$\ket{i}$ in the ground-state wave function when expanded in a {\it given} orthonormal basis $\{
\ket{i} \}$: $\ket{\Psi_0} = \sum_i a_i \ket{i}$. The ground-state is normalized, ensuring that $\sum_i p_i=1$. 

Most of the results will be presented for ground-states of celebrated models of quantum magnetism. We define first the $S=1/2$ XXZ model:
\be
\label{eq:xxz}
{\cal{H}}_{\rm XXZ}=\sum_{\langle i j\rangle}J_{ij} \left(S_{i}^{x}S_{j}^{x}+S_{i}^{y}S_{j}^{y}+\Delta S_{i}^{z}S_{j}^{z}\right). 
\ee
where $\Delta$ is the Ising anisotropy and the nearest-neighbor coupling constant $J_{ij}$ can
spatially vary. At $\Delta=1$, this is the Heisenberg model which has an enhanced $SU(2)$ symmetry. We also consider the  $S=1/2$ quantum Ising model 
\be
\label{eq:ising}
{\cal{H}}_{\rm Is.} = - \sum_{\langle i,j \rangle} \sigma_{i}^{z}\sigma_{j}^{z} -h \sum_i \sigma_{i}^{x}
\ee
in a transverse magnetic field $h$. We use standard notations with $\boldsymbol{\sigma}=2 \cdot {\bf S}$ being Pauli matrices.

The reader will of course recognize the resemblance of Eq.~\ref{eq:SR1}, \ref{eq:SR2} and \ref{eq:PS}
with the definitions of the R\'enyi and von Neumann {\it entanglement} entropies (EE) as well as of
the entanglement spectrum (ES), discussed at length in this volume. The two sets of quantities have
a clearly different physical interpretation: the SR entropies quantify localization in a given
Hilbert space basis while von Neumann-R\'enyi entropies quantify entanglement between parts of a system. At the formal level, the distinction is also simple: the SR entropies and the participation spectrum consider the {\it diagonal elements} of the (full) density matrix, while the entanglement entropies and spectrum are based on the {\it eigenvalues} of the {\it reduced} density matrix. Note that at a later stage, we will also consider the diagonal elements of the reduced density matrix. 

The SR entropies and PS have received less attention than EE and ES. Nevertheless recent works
have shown that they have several useful interests for characterizing condensed matter
ground-states~\cite{stephan_shannon_2009,stephan_renyi_2010,zaletel_logarithmic_2011,stephan_phase_2011,alcaraz_universal_2013,luitz_universal_2014,luitz_shannon_2014,stephan_shannon_2014}. This paper has two goals: first, we attempt to provide in Sec.~\ref{sec:review} an introduction to the field of SR entropies and PS in condensed matter by stating basic considerations (Sec.\ref{sec:basic}), reviewing previous results (Sec.~\ref{sec:previous}) and pinpointing the relation to EE and ES when there is one. In the second part, we present results on PS for 1d and 2d quantum spin systems, either defined by the full ground-state wave-function (Sec.~\ref{sec:full}) or by its restriction imposed by a spatial bipartition of the system (Sec.~\ref{sec:sub}). This will be useful to make a comparison with EE and ES of the same bipartition. In particular, this allows us to discuss the issue of entanglement Hamiltonians ~\cite{li_entanglement_2008,calabrese_entanglement_2008}.

\section{Shannon-R\'enyi entropies in condensed matter systems}
\label{sec:review}

The first natural question raised by the definition of Eq.~\ref{eq:SR1},\ref{eq:SR2} and
\ref{eq:PS}: how can such quantities reveal some intrinsic properties of the wave-function as they
are obviously basis-dependent? This is in contrast with EE and ES which do not depend on the basis
and for which a large set of work has demonstrated the usefulness in characterizing physical
properties of $\ket{\Psi_0}$. Quite recently, several works reviewed in Sec.~\ref{sec:previous} have
demonstrated some aspects of universality in the scaling properties (with system size) of SR
entropies. While the basis-dependence question probably cannot be solved or even addressed in full
generality, the independence of the universal part in the scaling can be demonstrated in some cases
as discussed in Sec.~\ref{sec:basis}. We also provide elements in Sec.~\ref{sec:computing} on how to
compute numerically SR entropies and PS for many-body ground-states, making comparison between
different methods as well as with computations of the related EE and ES.  

\subsection{First considerations on SR entropies}
\label{sec:basic}

\subsubsection{Scaling of SR entropies in many-body ground-states }

It is illustrative to consider first the scaling of SR entropies for a simple wave-function with uniform coefficients and a fixed number ${\cal P}$ of basis states: $\ket{\Psi_0}= \frac{1}{\sqrt{{\cal P}}} \sum_i \ket{i}$, which is simply given by $S_q = \ln ({\cal P})$, independent of $q$. For a many-body problem where the Hilbert space scales exponentially with the number $N$ of particles (e.g. $2^N$ for $N$ spins $1/2$), one naturally expects a {\it volume law}
\be
\label{eq:volume} 
S_q = a_q N+ \dots 
\ee
where $a_q$ are some bounded constants ($0 \leq a_q \leq \ln(2)$ for spins $1/2$), which in the generic case depend non-trivially on $q$ as well as on microscopic details contained in $\ket{\Psi_0}$. In other words, the ground-state wave-function occupies an exponential number of states in a given (simple) basis, and this is found to be the generic situation. 

Two remarks are in order when comparing to the scaling behavior of EE. First, Eq.~\ref{eq:volume}
clearly contrasts with the {\it area law}, the leading behavior in the scaling of EE in most
condensed-matter ground-states. Secondly, a long series of work (see
e.g.~\cite{holzhey_geometric_1994,kitaev_topological_2006,levin_detecting_2006,calabrese_entanglement_2004,metlitski_entanglement_2009,metlitski_entanglement_2011,kallin_anomalies_2011})
has established that the sub-leading behavior (beyond the area law) of EE has universal aspects for
different states of matter. While the state of the literature on SR entropies is not as rich, there
are several established results reviewed in Sec.~\ref{sec:previous} which indicate that sub-leading terms (represented by dots in Eq.~\ref{eq:volume}) also have universal features.

\subsubsection{Relation to entanglement entropies and spectra for Rokhsar-Kivelson wave-functions }

There is a class of ground-states, Rokshar-Kivelson (RK) wave-functions~\cite{rokhsar_superconductivity_1988},  where there is an exact relation between ES and PS (of a different wave-function). We refer the reader interested to a precise definition of RK wave-functions and their properties to Refs.~\cite{castelnovo_from_2005,rokhsar_superconductivity_1988}, and just mention that they are ground-states of local Hamiltonians with a stochastic matrix form. Several interesting models support points in their phase diagram which have RK ground-states wave-functions, such as quantum dimer models~\cite{rokhsar_superconductivity_1988,moessner_resonating_2001} or quantum vertex models~\cite{ardonne_topological_2004}. For RK wave-functions, it can be shown~\cite{stephan_shannon_2009,furukawa_topological_2007} that the ES for a bipartition of the system (which may need precise lattice adjustments in some cases) living in dimension $d$ is {\it exactly} the PS of {\it another wave-function} living in dimension $d-1$, expanded in a {\it fixed} basis. 
For instance, the ES of $2d$ quantum dimer (respectively six-vertex) models is the PS of the ground-state of a free-fermionic (respectively XXZ spin) $1d$ chain. Of course, if ES and PS are identical, so are EE and SR entropies for such wave-functions. 

This remarkable correspondence leads to the following remarks: first, the area-law of the EE in
dimension $d$ is now understood as a volume law for SR entropies in dimension $d-1$. Second, note
that the construction~\cite{stephan_shannon_2009} allowing this correspondence {\it fixes} the basis
in which to compute the PS for the $d-1$-dimensional system. Finally and most
importantly, all results (\eg on subleading universal terms) obtained on SR entropies for systems in
dimension $d-1$ can be directly translated for EE of RK wave-functions in dimension $d$. Note that
RK wave-functions are often found as finely-tuned multi-critical points (\eg for the square lattice quantum dimer model~\cite{rokhsar_superconductivity_1988,ardonne_topological_2004}), therefore the obtained scaling may not be generic. However, there are RK wave-functions which are representative of the phase, for instance the topological $\mathbb{Z}_2$ liquid in the quantum dimer model on the triangular lattice is well captured by a RK wave-function~\cite{moessner_resonating_2001}. In those latter cases, RK wave-functions offer a unique opportunity for (analytical) exact computations of a non-trivial ES. This also gives another nice incentive to study the PS. 

\subsubsection{Replica picture} 

The participation $p_i=| \langle i | \Psi_0 \rangle |^2$ can be interpreted in the transfer matrix
point of view, by taking the scalar product of the basis state $\ket{i}$ with the ground-state
projected out by application of $\exp(-\beta H)$ with $\beta \rightarrow \infty$ to an initial
random state $p_i = | \langle i | \exp(-\beta H) | \Psi_{\rm ini} \rangle|^2/ \tilde{Z} $, with the
normalization factor $\tilde{Z}=\sum_j   | \langle j| \exp(-\beta H) | \Psi_{\rm ini}
\rangle|^2$. As usual, the initial wave-function $ | \Psi_{\rm ini}
\rangle$ should have a non-zero overlap with the ground-state (it should {\it e.g.} have the same quantum numbers). Consider now $p_i^q$ for an integer $q\geq 2$: it can formally be seen as obtained from
$q$ different applications of  $\exp(-\beta H)$ (with $\beta \rightarrow \infty$) on $q$ different
copies of the system which are {\it forced to coincide} on the resulting ``boundary'' state
$\ket{i}$~\cite{fradkin_entanglement_2006}. This replica trick is at the root of many analytical
\cite{fradkin_entanglement_2006,zaletel_logarithmic_2011,hsu_universal_2010} (see the illustrative
``R\'enyi book'' picture of Ref.~\cite{stephan_renyi_2010} with $n=q$ in their notations) as well as
numerical calculations~\cite{luitz_universal_2014} of the R\'enyi entropy Eq.~\ref{eq:SR1} for integer
$q\geq 2$. 

The parameter $q$ morally plays the role of an inverse temperature in the R\'enyi entropies,  ``emphasizing'' the effect of coefficients of the wave-function (the hand-waving argument is the following: if the $p_i=\exp(-E_i)$ assume a Boltzmann form, then $q$ appears as an inverse temperature $\beta=1/T$).  We will indeed observe phase transitions as a function of $q$. One should therefore be very careful in extrapolating analytical results obtained from a replica calculation to $q=1$ to compute the Shannon entropy Eq.~\ref{eq:SR2} (in some cases this is indeed not possible).

\subsubsection{Special values of $q$}

Besides all integers, there are two other remarkable values:  $q=1/2$ and $q=\infty$. Taking $q=\infty$ simply selects the state(s) $\ket{i_{\rm max}}$ with the
largest coefficient, {\it i.e.} with the largest probability $p_{\rm max}=| \langle i_{\rm max} |
\Psi_0 \rangle|^2$ (or equivalently the lowest pseudo-energy $\epsilon_0 = -\ln (p_{\rm max})$ in the
PS). This state may have a multiplicity ${\cal D}$ for instance due to symmetries of the wave-function
$\ket{\Psi_0}$. While it only carries information on a single  (out of an exponential number)
coefficient of the wave-function, the scaling of $S_\infty$ with system size contains a lot of information on the physical system, as discussed below in Sec.~\ref{sec:previous}. On the other hand, $S_{q=1/2}=2 \log(\sum_i \sqrt{p_i})=2\log(\sum_i |a_i|)$ treats on equal footing all coefficients of the wave-function. These coefficients $a_i$ furthermore appear {\it linearly} in the expression of $S_{q=1/2}$, which allows to anticipate the connection with a scalar product used below (see also Sec.~\ref{sec:1d}).

The most probable state(s) is (are) often easy to find: for instance for the XXZ model
(Eq.~\ref{eq:xxz}) and considering the $\{S^z\}$ basis, we always found them to be the N\'eel states
$\ket{{\cal N}_A}=\ket{ \uparrow_{\cal A} \downarrow_{\cal B}  \uparrow_{\cal A} \downarrow_{\cal B}
... }, \ket{{\cal N}_B}=\ket{ \downarrow_{\cal A} \uparrow_{\cal B}  \downarrow_{\cal A}
\uparrow_{\cal B} ... }$ for antiferromagnetic interactions $J_{\langle ij \rangle}>0$ on
bipartite lattices (with sublattices ${\cal A}$,${\cal B}$) and the two fully polarized states $\ket{\up
\up \up ... }$ and $\ket{\dw \dw \dw ... }$ for ferromagnetic interactions $J_{\langle ij\rangle }<0$. For the quantum
Ising model (Eq.~\ref{eq:ising}), it is the polarized state in the field direction $\ket{\up \up \up
...}_x $ for any $h>0$ for the $\{\sigma^x\}$ basis, and the two ferromagnetic states $\ket{\up \up
\up ... }_z$ and $\ket{\dw \dw \dw ... }_z$ in the $\{\sigma^z\}$ basis. For these models (and many
others that satisfy the conditions of the Perron-Frobenius theorem in an appropriate basis) for
which $\ket{i_{\rm max}}$ is simply the polarized state\footnote{for the antiferromagnetic XXZ
model, one performs a unitary transformation where all spins on one sublattice are rotated by $\pi$ around the $z$-axis.}
$\ket{\up \up \up \dots}_{x/z}$ in either the $\{S^x\}$ or $\{S^z\}$ basis, we have a simple
relation~\cite{luitz_universal_2014} between the two entropies expressed in different bases: 
\be
\label{eq:Sdual}
S^{x/z}_{q=1/2}=N \ln{2} - S^{z/x}_{q=\infty},
\ee
since $\ket{\up \up \up ...}_{x/z}= 2^{-N/2} \sum_i \ket{i}_{z/x}$ for $S=1/2$ systems. For states
enjoying $SU(2)$ symmetries (such as the ground-states of the Heisenberg model), this puts an even stronger condition for entropies in the {\it same} basis $S^{x/z}_{q=1/2}=N \ln{2} - S^{x/z}_{q=\infty}$. 
Note that the SR entropies $S_{q=1/2}$ and $S_\infty$ also play special roles in the CFT description as they correspond to special boundary fixed ($\ket{\up \up \up ...}$) or free ($\sum_i \ket{i}$) states in the appropriate basis (see Sec.~\ref{sec:1d}).

\subsubsection{Multifractality } 

Up to a factor $\ln(2)$, the prefactors $a_q$ of the volume law Eq.~\ref{eq:volume} are equal to the
{\it fractal dimensions} of the set of probabilities $p_i$ \cite{mirlin_multifractality_2000,atas_multifractality_2012}. A non-linear dependence of $a_q$ on $q$ signals multifractality, a phenomenon that arises in various different complex systems. For instance, it is a key feature of the Anderson transition~\cite{evers_anderson_2008,evers_fluctuations_2000}. Given that the SR entropies quantify localization in a (Hilbert space) basis, it is not unexpected that multifractality is also found in its scaling behavior. What is clearly different however from single-particle problems (such as the Anderson localization) is that multifractality is {\it always} present in many-body systems (except for a few isolated cases). This was first remarked in Ref.~\cite{atas_multifractality_2012} in spin chains, and further confirmed in Ref.~\cite{luitz_universal_2014} on the basis of quantum Monte Carlo simulations. In fact, multifractality can be exactly shown to occur even for simple featureless wave-functions (such as the isolated plaquette limit of the model studied in Ref.~\cite{luitz_shannon_2014}). This indicates that fractal dimensions $a_q$ are not good quantities to characterize the nature of the phase in many-body ground state wave functions, in contrast to single-particle physics where they can distinguish metallic, insulating and critical phases. As discussed in Sec.~\ref{sec:previous}, subleading terms of Eq.~\ref{eq:volume} play this role for many-body systems.

Let us mention that, even though the value of $a_q$ does not contain physical information, its {\it
variation} with a parameter in the model (such as the transverse field $h$ in Eq.~\ref{eq:ising})
may capture phase transitions (see the inflection points signaled in
Ref.~\cite{luitz_universal_2014,luitz_shannon_2014}).

\subsection{Universal subleading terms}
\label{sec:previous}

We review here several results where the subleading terms in the scaling of SR entropies has been established either analytically (mainly in $d=1$, either exactly or with the use of CFT) or numerically (using methods reviewed in Sec.~\ref{sec:computing}). 

\subsubsection{One-dimensional systems}
\label{sec:1d} 

\paragraph{Luttinger liquids }
    
    Let us start with the probably best understood case of Luttinger Liquid (LL)
systems, {\it i.e.} which can be described by a massless free boson field theory in dimension $1+1$,
with a central charge $c=1$. This is the case for instance of the critical phase of the XXZ spin
chain (for $-1 < \Delta \leq 1$). Thanks to specific properties of the free-boson field theory,
calculations can be made~\cite{stephan_shannon_2009} without resorting to replica formulations (for a replica formulation, see Ref.~\cite{oshikawa_boundary_2010}). In the basis where the conserved
charge is diagonal (namely the $\{S^z\}$ basis for the XXZ spin chain), the following scalings for the SR entropies for a chain of size $L$ (with $N=L$ spins 1/2) with periodic boundary conditions (PBC) are obtained~\cite{stephan_shannon_2009}:
\be
\label{eq:bscaling}
S_q = a_q N + b_q + o(1)
\ee
with the subleading constant term experiencing a phase transition as a function of $q$
\begin{eqnarray}
\label{eq:bLL1}
b_q=-\frac{1}{2}\left(\ln{K} + \frac{\ln{q}}{q-1}\right)  &\quad {\rm for} \quad & q \leq q_c = K {\cal D}^2 \\
b_q=\frac{1}{1-q}\left( q \ln \sqrt{K} + \ln{\cal D}\right) &\quad{\rm for} \quad& q \geq q_c 
\label{eq:bLL2}
\end{eqnarray}
with $K$ the LL parameter, and ${\cal D}$ is the multiplicity of $\ket{i_{\rm max}}$, the most probable state.
For the XXZ spin chain, the LL parameter is given by $K=(2-2\arccos(\Delta)/\pi)^{-1}$ and ${\cal
D}=2$. The predictions Eqs.~\ref{eq:bLL1} and \ref{eq:bLL2} have been checked numerically in the
XXZ spin chain \cite{stephan_shannon_2009} and for free-fermions models \cite{stephan_shannon_2009} (allowing high-precision numerics and in some cases exact results), as well as for a spin ladder in a magnetic field~\cite{luitz_universal_2014}. 

In the free boson field theory, the values of $S_q$ at $q=1/2$ and $q=\infty$ are special as they
are related in the transfer-matrix approach to partition functions of an infinite half-cylinder with respectively free or fixed boundary conditions~\cite{stephan_shannon_2009}, which are conformally invariant. This relates $b_{1/2}$ and $b_\infty$ to Affleck-Ludwig boundary entropies~\cite{affleck_universal_1991}, well-known in other contexts described by CFT. Note that closed formulas for $S_\infty$ have been recently obtained for the XXZ spin chain using Bethe ansatz~\cite{pozsgay_overlaps_2013,brockmann_generalized_2014}.

What is remarkable is that the SR entropies give direct access, with a good precision and relatively modest computational effort, to the LL parameter $K$, which is often hard to estimate numerically (see however recent efforts in Ref.~\cite{Rachel12,Lauchli13}). This is particularly interesting in the case of $S_\infty$ which boils down to calculating the overlap of the ground-state wave-function with a single state (this was already remarked in calculations of fidelity~\cite{camposvenuti_universal_2009}). Different lattice models with the same value of $K$ will have different $a_q$ leading terms, but share the same subleading $b_q$ term for $q \leq q_c$ (in the regime $q\geq q_c$, the degeneracy ${\cal D}$ which may be different for different models also enters the formula). 

For chains with {\it open}  boundary conditions (OBC), the SR entropies acquire a {\it logarithmic} subleading term:
\be
\label{eq:logscaling}
S_q = a_q N + l_q \log(N) + \tilde{b}_q + o(1)
\ee
which is also universal and takes the following values~\cite{zaletel_logarithmic_2011,stephan_phase_2011}
\begin{eqnarray}
\label{eq:lLL1}
l_q= - \frac{1}{4}  \quad &  {\rm for} & \quad \!\!    q < q_c = K {\cal D}^2 \\
l_q= \frac{q}{q-1}\left( \frac{1}{4K} -\frac{1}{4} \right) \quad &  {\rm for} & \quad\!\!   q > q_c.
\label{eq:lLL2}
\end{eqnarray}
The value at $q=q_c$ is not known in general, but was found numerically in the specific case of the
XX model (Eq.~\ref{eq:xxz} at $\Delta=0$ for which $K=1$) to be~\cite{stephan_phase_2011} $l_{q_c=4}=-\frac{1}{6}$. Note that the next sub-leading term (denoted $\tilde{b}_q$ to avoid confusion with $b_q$ in Eqs.~\ref{eq:bLL1} and \ref{eq:bLL2}) has no reason to be universal.

The existence of a log term in Eq.~\ref{eq:logscaling} is understood as a corner
contribution~\cite{cardy_finite_1988} to the free energy in the CFT formulation of the problem. When $q>q_c$, there is also a contribution from
boundary conditions changing operators present due to the OBC (here, this is the factor $\frac{1}{4K}$ for OBC -- different $l_q$ would be obtained for
different boundary conditions). The phase transition observed in the universal subleading
coefficients (Eqs.~\ref{eq:bLL1}, \ref{eq:bLL2}, \ref{eq:lLL1}, \ref{eq:lLL2}) is
understood~\cite{stephan_shannon_2009,stephan_phase_2011} as a {\it boundary} roughening transition caused by
vertex operators present for lattice models: the free field gets locked onto ${\cal D}$
configurations corresponding on the lattice to the most probable states $\ket{i_{\rm max}}$.
Correspondingly, the subleading terms of the entropies are dominated by the contributions of these
states, which take the values $\frac{1}{1-q}\log({\cal D} p_{\rm max}^q)$. This phase transition would be missed by a straightforward replica analysis.

Finally, let us mention that when $\Delta>1$, the ground-state of the XXZ spin chain does no longer realize a Luttinger Liquid, but rather adopts Ising antiferromagnetic order, spontaneously breaking the $\mathbb{Z}_2$ spin inversion symmetry. In this case of a discrete symmetry breaking with a two-fold degenerate ground-state, the subleading terms of the SR entropies are also universal, taking the form of Eq.~\ref{eq:bscaling} with $b_q=\ln(2)$ for all $q$~\cite{stephan_shannon_2009}.

\paragraph{Quantum Ising chain } 

SR entropies for the quantum Ising chain (Eq.~\ref{eq:ising} in $d=1$) have also been studied in
detail, in particular at the critical point thanks to CFT and high-precision numerics. In $1d$, the quantum Ising chain has a quantum phase transition at $h_c=1$ between a low-field ferromagnetic phase and a high-field polarized (paramagnetic) phase. The phase transition is in the $2d$ Ising universality class, described by a  $c=1/2$ CFT different from the $c=1$ free boson theory mentioned earlier. This will lead to quite different contributions to the universal subleading terms of the SR entropies.

Looking at Hamiltonian Eq.~\ref{eq:ising}, there are two ``natural'' choices for bases in which to
compute SR entropies: the $\{\sigma^z\}$ and $\{\sigma^x\}$ basis. For the $1d$ quantum Ising chain, the corresponding SR entropies are related by a Kramers-Wannier duality~\cite{stephan_shannon_2009}: $S^z_q(h)=S^x_q(1/h)+\ln(2)$. For chains with PBC, the scaling of entropies is again captured by Eq.~\ref{eq:bscaling} with the universal constant term taking the values~\cite{stephan_shannon_2009} reported in Table~\ref{tab:b}

\begin{table}[!h]
\centering
\begin{tabular}{l|c|c|r }
& $b_q^x$ & $b_q^z$ & Regime\\
\hline
\hline
$h<h_c, \forall q$ & $-\ln 2 $ & $\ln 2$ & Ordered\\
\hline
$h>h_c, \forall q$ & $0$ & $0$ & Disordered\\
\hline
$h=h_c, q<q_c=1$ & $-\ln 2$ & $0$ & Critical\\
\hline
$h=h_c, q>q_c=1$ & $0$ & $\ln 2$ & Critical \\
\hline
$h=h_c, q=q_c=1$ & $b_1^*-\ln 2$ & $b_1^*$ & Critical \\
\end{tabular}
\caption{Subleading constant terms of the SR entropies for the transverse field Ising chain.}
\label{tab:b}
\end{table}

Away from criticality, the subleading constants are easily understood by simply considering the limits $h=0$ and $h\rightarrow \infty$ respectively. 
At criticality, the subleading term takes non-trivial values, with a phase transition as a function of $q$. For $q>q_c=1$, the subleading term is dominated (``attracted") by its value for $q=\infty$, corresponding to the ${\cal D}=2$ degenerate most probable ferromagnetic states $\ket{\up \up \up ...}_z ,\ket{\dw \dw \dw ...}_z$ in the $\{\sigma^z\}$ basis and the non-degenerate polarized state $| \up \up \up ... \rangle_x$ in the in the $\{\sigma^x\}$ basis. For $q<q_c$, the subleading term is attracted by its value at $q=1/2$, which can be related (see Eq.~\ref{eq:Sdual}) to the values at $q=\infty$ in the other basis: $b_{q=1/2}^{(x/z)}=-b_{q=\infty}^{(z/x)}$.
In the $\{\sigma^z\}$ basis (which is the ``good" basis for the CFT), the value at $q=1/2$ (respectively  $q=\infty$) is again understood~\cite{stephan_shannon_2009} in terms of Ludwig-Affleck boundary entropies for the free (respectively fixed) conformally invariant boundary conditions,  which act as ``attractive" fixed points.

On the other hand, at $q=1$ the subleading Shannon constant $b_1$ takes a non-trivial value at the quantum phase transition estimated with high numerical precision~\cite{stephan_renyi_2010} to be $b^*_1= 0.2543925(5)$ (an improved analysis~\cite{lau_information_2012} gives $b^*_1= 0.254392505(10)$). This constant appears in the subleading term of the Shannon entropy for several different models in the 2d Ising universality class (and is therefore universal), but it is still not understood at the moment from CFT. 

The phase transition taking place at $q_c=1$ for the quantum Ising model would again be missed by a replica formulation of the SR entropies, for which analytical continuations to $q=1$ would predict wrong subleading terms in this case.

For OBC, a logarithmic subleading term arises at the critical point, with SR entropies scaling as Eq.~\ref{eq:logscaling} with~\cite{zaletel_logarithmic_2011,stephan_phd_2011}:
\begin{eqnarray}
l_q^x(h_c)= \frac{3q}{8(q-1)} \quad & \quad l_q^z(h_c) =  -\frac{q}{8(q-1)} \quad &  \quad q < q_c=1 \\
l_q^x(h_c)=  -\frac{q}{8(q-1)} \quad & \quad l_q^z(h_c) =  \frac{3q}{8(q-1)} \quad &  \quad q > q_c=1
\end{eqnarray}
The behavior at the $q_c=1$ phase transition is special, as there $l_1$ itself diverges logarithmically~\cite{stephan_shannon_2014} with $N$, {\it i.e.} the first subleading term is ${\ell}_1^x (\ln{N})^2$ with ${\ell}_1^x=-0.02934(5)$~\cite{stephan_shannon_2014}. 

Once again, the values of the subleading logarithmic coefficient can be understood from CFT~\cite{zaletel_logarithmic_2011} in the limits $q>q_c$ ($q<q_c$), which are attracted by the fixed and free conformal invariant boundary conditions at $q=\infty$ and $q=1/2$ respectively. 

\subsubsection{Two dimensional quantum systems}
\label{sec:2d}

The results presented in the previous section clearly demonstrate that the powerful techniques available in $1d$ (CFT for critical systems, exact diagonalization, free-fermionic exactly solvable points etc) are extremely useful to determine the scaling behavior of SR entropies. They are no longer available for $2d$ quantum systems, nevertheless similar investigations can be performed using quantum Monte Carlo (QMC) techniques~\cite{luitz_universal_2014,luitz_shannon_2014}.  

\paragraph{Quantum Ising model } 

In $2d$, the quantum Ising model (Eq.~\ref{eq:ising}) also exhibits a continuous quantum phase transition between a low-field ferromagnetic and a high-field paramagnetic phase, which belongs to the $3d$ Ising universality class. The QMC simulations of Ref.~\cite{luitz_universal_2014} support for systems with PBC a SR entropy scaling similar as in the 1d case (Eq.~\ref{eq:bscaling}) (with $N=L^2$ spins, $L$ being the linear dimension), with identical values for $b_q^x$ and $b_q^z$ in the ordered and disordered phases (Table~\ref{tab:b}). At the critical point $h_c$, the behavior is different. At $h_c$, only simulations in the $\{\sigma^x\}$ basis and for $q=2,3,4,\infty$ allow to reach large enough system sizes such as to probe the scaling behavior. The results are consistent with a scaling 
\be 
\label{eq:b2d}
b_{q\geq 2}^x(h_c)=\frac{q}{q-1}b_\infty^x (h_c)
\ee
 with $b_\infty^x (h_c)=0.19(1)$, suggesting that again the (non-degenerate) most probable state $|\up \up \up ... \rangle_x$ dominates the subleading scaling in this regime. On the other hand, results~\cite{unpublished} in the $\{\sigma^z\}$ basis for $q=\infty$ also allow (see Eq.~\ref{eq:Sdual}) to obtain $b_{q=1/2}^x=-\ln(2)$, suggesting a $q$-induced phase transition at a value $1/2 \leq q_c<2$.  The universality was ascertained by considering the quantum phase transitions in two different models (on the square and triangular lattices), indicating that the results in Eq.~\ref{eq:b2d} should also be characteristic of the 3d Ising universality class. Note that direct simulations at the critical point of the 3d Ising model  would allow to access $b^z_q(h_c)$ ({\it i.e.} the similar constants in the $\{\sigma^z\}$ basis).  

\paragraph{XXZ model } 

In two dimensions, one can study the signature of zero temperature spontaneous breaking of {\it continuous} symmetries in the behavior of subleading terms in SR
entropies. The XXZ model (Eq.~\ref{eq:xxz}) on the square lattice serves well this
purpose, as the ground-state breaks $U(1)$ symmetry when $-1 < \Delta < 1$, $SU(2)$ symmetry for
$\Delta=1$ and a discrete $\mathbb{Z}_2$ Ising symmetry for $\Delta > 1$. 

For the Heisenberg model ($\Delta=1$), simulations for $q=2,3,4,\infty$ on the square lattice indicate that the scaling Eq.~\ref{eq:logscaling} holds~\cite{luitz_universal_2014}. Precise determination of $l_q$ are difficult with the current sizes at hand since they result from a fit of a logarithmic subleading term on about 2 decades in $N$. A value $l_\infty \sim 0.5-0.6$ has been obtained for the Heisenberg model on the square lattice  and estimates for $l_2,l_3$ and $l_4$ indicate that a scaling $l_{q \geq 2}=\frac{q}{q-1}l_\infty$ could hold. We do not specify the $\{S^x\}$ or $\{S^z\}$ basis here as they are equivalent under a SU($2$) rotation. On the other hand, exact results~\cite{luitz_universal_2014} on the Lieb-Mattis model~\cite{lieb_ordering_1962} (a toy model for SU($2$) symmetry breaking) give $l_{q > 1}=\frac{q}{q-1}l_\infty$ with $l_\infty=1$ (as well as $l_1=0$ and $l_{0<q<1}= \frac{-1}{2(1-q)}$). The combination of these two results suggest the potential importance of spin-waves (Goldstone modes) in these subleading terms, as they are {\it not} present in the Lieb-Mattis model. 

For the XXZ model in the $U(1)$-symmetry breaking regime ($0 \le\Delta <1$), large-scale simulations are only possible in the $\{S^x\}$ basis, and a similar scaling~\cite{luitz_universal_2014,unpublished} is found $l_{q > 1}^x=\frac{q}{q-1}l_\infty^x$ with $l_\infty^x \sim 0.25-0.3 $ (again with the same precaution on the available system sizes). On the other hand, in the Ising gapped regime $\Delta > 1$, the logarithmic subleading terms vanishes and one recovers the scaling of Eq.~\ref{eq:bscaling} with $b_q^z=\ln(2)$, as understood from the $\Delta \rightarrow \infty$ limit of this phase which breaks a $\mathbb{Z}_2$ discrete symmetry.

\paragraph{Quantum phase transition in a SU($2$) model } 

The subleading terms in the R\'enyi entropy $S_\infty$ have been studied in
Ref.~\cite{luitz_shannon_2014} for the quantum phase transition between an antiferromagnet (breaking
SU($2$) symmetry) and a paramagnetic phase (with a single non-degenerate ground-state) for an
Heisenberg model with two spatially varying coupling constants $J_{\langle i,j \rangle}=J_1,J_2$,
with ratio $g=J_2/J_1$. Logarithmic corrections with a presumably constant prefactor $l_\infty$ as a
function of $g$ also appear in the full antiferromagnetic phase, whereas they vanish in the quantum
disordered phase. Due to the large value of $a_\infty$ in this phase, the QMC simulations of
Ref.~\cite{luitz_shannon_2014} could not exactly pinpoint the value of the resulting $b_\infty$,
even though a value $b_\infty=0$ is expected. At the quantum phase transition $g_c$, the same caveat
applies, but results for a line subsystem (see below Sec.~\ref{subsec:subentropy}) suggest that possibly a universal value $b_\infty^*(g_c)$, characteristic of the 3d $O(3)$ universality class to which the quantum phase transition belongs, could emerge.

\subsubsection{SR entropies of sub-sytems}
\label{subsec:subentropy}

The SR entropies can also be defined and studied for restrictions of the wave-function, most
illustratively for a spatial bipartition of the full system ({\it subsystems}). Eqs.~\ref{eq:SR1}
,\ref{eq:SR2} and \ref{eq:PS} are similar except that now $p_i=\rho^A_{ii}$, {\it i.e.} the diagonal entry of the {\it reduced} density matrix on the subsystem $A$ ($\rho^A = \Tr_B \ket{\Psi_0} \bra{\Psi_0}$ where $B$ is the spatial complement of $A$). 

For a subsystem $A$ containing $x$ spins $1/2$, we naturally expect a volume scaling $S_q^A= a_q^A x + ... \quad \!\!\!$ for the SR entropies. Being interested in subleading terms, it is natural to consider the SR {\it mutual information}
\be
\label{eq:mutual}
I_q = S_q^A+S_q^B - S_q^{A\cup B}
\ee
where $S_q^{A\cup B}$ is the SR entropy of the full system, as studied in previous sections. This linear combination will only pick up the subleading terms (volume terms cancel).

\paragraph{Subsystems in 1d critical chains } 

For a chain subsystem of size $x$ embedded in a critical spin chain of size $L$ with PBC, several recent works considered the following scaling~\cite{um_entanglement_2012,lau_information_2012,alcaraz_universal_2013,stephan_emptiness_2013}:
\be
\label{eq:mutualscaling}
I_q = \frac{\gamma_q}{4} \log(\tilde{x}) + ...
\ee
where $\tilde{x}=\frac{L}{\pi}\sin (\pi x / L)$ is the conformal distance and where, based on numerics, $\gamma_1$ was conjectured~\cite{alcaraz_universal_2013} to be the central charge $c$ of the CFT describing the critical spin chain. Note the similarity with the celebrated formula for EE for 1d critical spin chains~\cite{calabrese_entanglement_2004}. In fact, it can be shown~\cite{stephan_shannon_2014} for the free boson theory (valid for LL) with $c=1$ that Eq.~\ref{eq:mutualscaling} is correct with $\gamma_q=c$ up to the critical value $q_c=K{\cal D}^2$ mentioned in Sec.~\ref{sec:1d}. For larger $q$, the physics is once again dominated by the most probable states, and the result $\gamma_q=\frac{cq}{(q-1)}$ is found~\cite{stephan_shannon_2014}. For the critical Ising chain (Eq.~\ref{eq:ising}) with $c=1/2$, results of Ref.~\cite{stephan_emptiness_2013} indicate $\gamma_{q>1}=\frac{cq}{(q-1)}$. The case $q=1$ is again special for the Ising chain, and precise numerical simulations~\cite{stephan_shannon_2014} prove the conjecture of Ref.~\cite{alcaraz_universal_2013} wrong and find $\gamma_1=0.4801629(2)\neq c = 1/2$ (earlier numerical simulations found $\gamma_1=0.49$~\cite{alcaraz_universal_2013}, $\gamma_1=0.476$~\cite{um_entanglement_2012}, $\gamma_1=0.4804(8)$~\cite{lau_information_2012} and $\gamma_1=0.4802(4)$~\cite{stephan_phd_2011}). 

When the critical chain possesses OBC, a scaling similar to Eq.~\ref{eq:mutualscaling} is obtained for the free boson theory, albeit with correspondingly modified values of $\gamma_q$~\cite{stephan_emptiness_2013,stephan_shannon_2014}. Note that Ref.~\cite{stephan_emptiness_2013,stephan_shannon_2014} also predicts unusual further subleading terms (dots in Eq.~\ref{eq:mutualscaling}) in some other cases.

\paragraph{Line subsystems in 2d systems }

All the results in the previous paragraph were obtained thanks to exact numerical (see
Sec.~\ref{sec:computing}) or CFT methods only available in 1d. QMC methods allow to treat subsystems
in $d>1$ systems, and Ref.~\cite{luitz_shannon_2014} studied the SR entropy $S_\infty$ for a
subsystem composed of a {\it line} of linear size $L$ in a sample containing $L\times L$ spins with
PBC. The advantage of a line subsystem is that much larger systems can be studied, as the volume
term only scales with $L$, and not as $L^2$ for the full system. For the quantum phase transition in
a $S=1/2$ Heisenberg system with varying couplings, Ref.~\cite{luitz_shannon_2014} gives evidence for a scaling 
$$S_\infty^{\rm line} = a_\infty^{\rm line} L + l_{\infty}^{\rm line} \log(L)+...$$
in the antiferromagnetic phase (with continuous symmetry breaking), remarkably mimicking the result
for the SR entropy $S_\infty$ in the full system (Eq.~\ref{eq:logscaling}). In the quantum
disordered phase and at the quantum critical point, the subleading logarithmic term disappears, giving rise to:
$$S_\infty^{\rm line} = a_\infty^{\rm line} L + b_{\infty}^{\rm line}+...$$
with  $b_{\infty}^{\rm line}=0$ in the quantum disordered phase, and $b_{\infty}^{\rm
line}(g_c)=0.41(1)$ at the quantum critical point. Numerical evidence for $b_{\infty}^{\rm
line}(g_c)$ being a universal constant for the $3d$ $O(3)$ universality class is provided by QMC
simulations of $b_{\infty}^{\rm line}(g_c)$ for two quantum phase transitions in $2d$ as well as for a $3d$ finite-temperature phase transition belonging to this universality class~\cite{luitz_shannon_2014}.

\subsubsection{Other 2d quantum wave-functions } 

The SR entropies of subsystems and the related SR mutual information have also been calculated for
model wave-functions, mostly of RK type and for geometries where the subsystem is half the full
system. Note again that this is different from the EE calculation mentioned in Sec.~\ref{sec:basic}.
The classical-quantum correspondence built in the RK wave-functions ensures that this is equivalent
to computing SR entropies (and mutual information) in a classical problem, with a focus on computing universal subleading scaling terms (such as {\it e.g.} the ``classical topological entropy"~\cite{castelnovo_topological_2007,hermanns_renyi_2013}). We refer the interested reader to the related Refs.~\cite{castelnovo_topological_2007,wilms_mutual_2011,lau_information_2012,iaconis_detecting_2013,rahmani_universal_2013, hermanns_renyi_2013,stephan_geometric_2014} for more details.

\subsection{Computing SR entropies and participation spectra}
\label{sec:computing}

\subsubsection{Analytical calculations } 

Computing the SR entropies and PS exactly is sometimes possible analytically for some specific
models, such as free fermions or RK wave-functions. Besides these cases, CFT has also proven
extremely useful to derive scaling of SR entropies for $1d$ critical systems  (see
Sec.~\ref{sec:previous} where we reviewed most of the previous literature). However, in general much
less analytical results are available for SR entropies, in contrast to EE and ES (see {\it e.g.}
Refs.~\cite{holzhey_geometric_1994,calabrese_entanglement_2004,calabrese_entanglement_2008,levin_detecting_2006,kitaev_topological_2006,Lauchli13}
amongst many other important analytical contributions). The main reason is probably not due to an
additional technical complexity (actually the calculation of SR entropies is certainly more
tractable in many ``natural" bases), but most certainly because the SR entropies and PS have attracted less interest. Note however that while the exact calculation of EE is much simplified for free-fermion or free-boson models~\cite{peschel_reduced_2009}, this is {\it not} the case for SR entropies which remains an exponential problem. Field theory calculations for SR entropies subleading terms, similar to those available for EE~(see {\it e.g.} Refs.~\cite{casini_entanglement_2009,metlitski_entanglement_2009,furukawa_entanglement_2011,metlitski_entanglement_2011,klebanov_renyi_2012}), would be particularly welcomed.

\subsubsection{Numerical methods } On the other hand, computing SR entropies and PS with numerical
methods is probably simpler  than for EE and ES. Computing the PS with exact diagonalization
techniques where the ground-state wave-function is calculated exactly is straightforward. This has
been used extensively (mainly for the calculations of SR entropies), it is however limited to about
$N=40$ spins $1/2$. Besides the cost of the exact diagonalization (exponential in $N$), the simple
need of enumerating the $2^N$ states for calculating the full PS is also a limitation (one can of course use all symmetries available to reduce this enumeration). This
restriction will also be true for other methods which determine ``quasi-exactly'' the ground-state: density-matrix renormalization group (DMRG) or time-evolving block decimation (TEBD) as used in Ref.~\cite{zaletel_logarithmic_2011}. Note that DMRG, which is usually the method of choice to compute the ES, can indeed be used to compute the lowest-lying part of the PS as well~\cite{McCulloch_unpublished}. 

The constraint of the total size of the PS can be relieved if one considers {\it sampling}, instead of completely computing, the set of  $p_i$. As demonstrated in Ref.~\cite{luitz_universal_2014}, this is the full strength brought by quantum Monte Carlo methods: indeed, a state $\ket{i}$ of the computational basis is seen {\it exactly} in proportion to $p_i$ in QMC. One just needs to record the histogram of $\ket{i}$ observed in the QMC simulations to estimate the PS. Note that  only the PS and SR entropies in the {\it QMC computational basis} ({\it i.e.} where the projector $\ket{i}\bra{i}$ is diagonal) are readily obtained. For full system PS, QMC techniques can reach typically the same number of spins $N$ as in exact diagonalization, and even slightly larger numbers in the cases where the wave-function is well-localized in the computational basis. For subsystems on the other hand,  QMC can go much further, the main limitation being the size of the subsystem itself. From the PS, all SR entropies for all $q$ are available, including the Shannon entropy $S_1$. Quite importantly, the SR entropies Eq.~\ref{eq:SR1} for integer $q\geq 2$ can be computed (without using histograms) by a simple adaptation~\cite{luitz_universal_2014} of a replica trick (see Sec.~\ref{sec:basic}): $q$ independent copies of the system are simulated in parallel, and a contribution to $S_q$ is obtained when the same state $\ket{i}$ is observed simultaneously on the $q$ copies. Additionally, the SR entropy $S_\infty=-\ln( p_{\rm max}) $ is also very easily obtained by recording the frequency of observation of the most probable state(s) $\ket{i_{\rm max}}$.

Overall, QMC simulations allow to compute SR entropies of large systems (up to $N=500$ spins $1/2$
for some systems) with a very good accuracy. If the entropies are too large however, the stochastic
sampling is not efficient anymore: a good rule-of-thumb is that entropies $S  \gtrsim 20$ (corresponding
to events with probabilities $< 2\cdot 10^{-9}$ ) cannot be reached with histogram based Monte Carlo sampling.
Finally, let us recall that the QMC sampling of SR entropies and PS is possible only when the QMC method is efficient ({\it i.e.} for models with no sign problem). Such methods compare favorably with respect to QMC methods for studying entanglement entropies: there only Renyi EE for integer $q\geq 2$ are available~\cite{hastings_measuring_2010,humeniuk_quantum_2012} while the direct sampling of the single-copy entanglement $S_{q=\infty}$ is out of reach. We note that recent developments allow the computation of the ES for fermionic models amenable to auxiliary field QMC~\cite{grover_entanglement_2013,assaad_entanglement_2013}, as well as the low-lying levels of the ES in the general case~\cite{chung_entanglement_2013}. 

Finally, we expect that other  traditional numerical methods in condensed matter many-body physics (such as series expansion) could be adapted  to the computation of SR entropies, even though this has not been the case until now.

\subsection{Basis dependence:}
\label{sec:basis}

We finally return to the question of the basis dependence of SR entropies and PS.  First, let us mention that there are cases where a specific basis is singled out exactly, for instance through the entanglement spectrum construction of RK wave-functions in dimension $d+1$ (see Sec.~\ref{sec:basic}). In other cases, specific bases are singled out in practice (``natural bases''), as {\it e.g.} the eigenbasis of the operators which constitute the Hamiltonian ($\{\sigma^x\}$ and $\{\sigma^z\}$ basis for the quantum Ising model Eq.~\ref{eq:ising}) or the computational basis in QMC simulations. 

We restate that there is no generic proof that the form of subleading terms discussed in Sec.~\ref{sec:previous} are independent of the basis, but  some further interesting remarks can be made. One should first remark that the separation of the leading (extensive) part from the subleading one makes sense only if the basis has some notion of locality (such as by using tensor products of local spin configurations). Given this, we note that there are some relations between results in different bases, such as the ``duality" relation Eq.~\ref{eq:Sdual} or the Kramers-Wannier duality~\cite{stephan_shannon_2009} for the 1d quantum Ising model, which strongly constraint the subleading terms. Also, there are several analytical or numerical studies~\cite{stephan_shannon_2009,stephan_renyi_2010,alcaraz_universal_2013,luitz_universal_2014} where the relation between subleading terms in different bases has been established for specific models. St\'ephan further established~\cite{stephan_phd_2011} for the $1d$ quantum Ising model, that any local rotation of the basis (between  the $\{\sigma^x\}$ and $\{\sigma^z\}$ basis) preserves the universal character of the subleading term, including the highly non-trivial value for $b_1$ taken at the quantum critical point. Finally, a very strong constraint is provided for $1d$ quantum critical systems by boundary CFT, as it is understood~\cite{stephan_shannon_2009,stephan_renyi_2010,stephan_phd_2011} that there are only a limited number of values for the subleading corrections, each of which corresponds to a different conformally invariant boundary condition which are of limited number. For instance there are only two types of unequivalent conformally invariant boundary conditions for the Ising universality class: free or fixed (picked by the $\{\sigma^x\}$ and $\{\sigma^z\}$ basis), and it is therefore expected that calculating the SR entropy in any (local) basis will inexorably lead to one of the two cases. 

Several results reviewed in Sec.~\ref{sec:previous} were obtained on the behavior of the SR entropies, but very few works (with the notable exceptions of Refs.~\cite{stephan_shannon_2009,stephan_renyi_2012,luitz_shannon_2014}) provide quantitative results on the PS. We attempt to fill this gap by considering general analytical arguments as well as numerical results for the PS of the full system in Sec.~\ref{sec:full}. We also consider the PS of a line subsystem in a $2d$ quantum system in Sec.~\ref{sec:sub}, making connection to the ES of this subsystem.

\section{Full participation spectra}
\label{sec:full}

In this section, we focus the discussion on participation spectra of ground state wave functions for
the full system without any bipartition.

\subsection{Analytical result for the participation gap: consequences for the entanglement gap}
\label{sec:gap}

One can exactly access to the first gap in the PS for a large class of short-range interacting models, using a simple argument. Let us illustrate this for the XXZ model  Eq.~\ref{eq:xxz} on a bipartite lattice in $d$ dimensions (having $N$ sites and $N_b$ bonds). We decompose ${\cal H}_{\rm xxz}$ in its diagonal part (in the $\{S^z\}$ basis) ${\cal H}_{d}=J_z \sum_{\langle i j\rangle} S_{i}^{z}S_{j}^{z}$ and its off-diagonal part ${\cal H}_{od}=J_{{xy}}\sum_{\langle i j\rangle}\left(S_{i}^{x}S_{j}^{x}+S_{i}^{y}S_{j}^{y}\right)$

\begin{figure}[t]
\bc
\includegraphics[width=0.6\columnwidth]{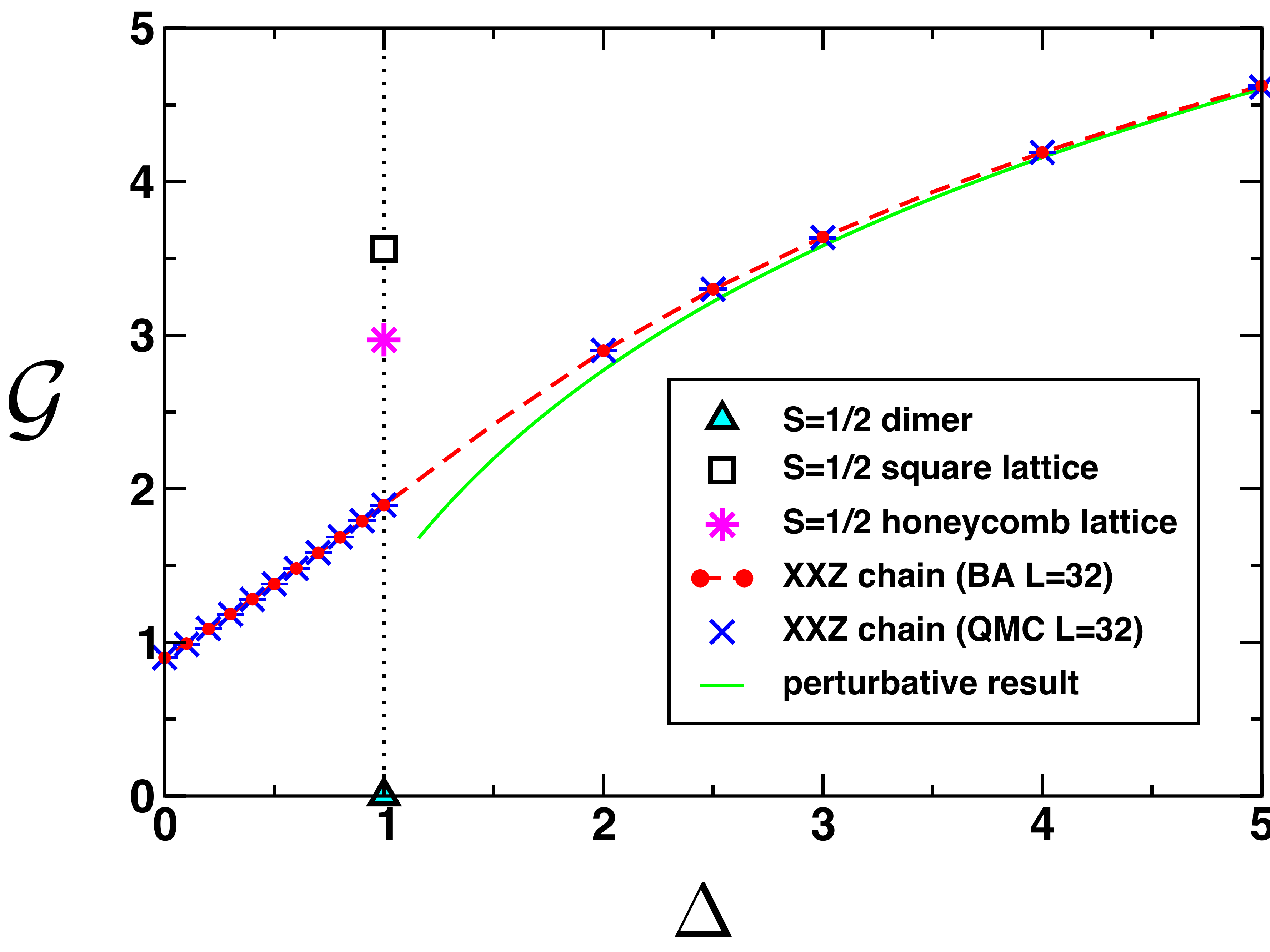}
\caption{First gap $\cal G$ in the full participation spectrum for various model
antiferromagnets. The (red) dashed curve displays the analytical result for $\mathcal{G}$ obtained
from the Bethe ansatz solution for $E_0$~\cite{Bethe31} and Eq. \ref{eq:exactG}, which agrees perfectly with the QMC
result for $L=32$ sites. The (green) continuous curve is the $1/\Delta$ perturbative result of Sec.~\ref{sec:xxzpert}. Values for an isolated $S=1/2$ dimer, and two-dimensional Heisenberg $S=1/2$ models (in the thermodynamic limit) on square and honeycomb lattice~\cite{Sandvik97,fouet_investigation_2001} are also shown.}
\ec
\label{fig:Gaps}
\end{figure}

The ground-state $|\Psi_0\rangle$ can be written in the following manner
\be
|\Psi_0\rangle=a_{\rm max}\Bigl(|{\cal N}_A\rangle+|{\cal N}_B\rangle\Bigr)
+a'_{\rm max}\sum_{k}\Bigl(|\varphi_{A,k}\rangle+|\varphi_{B,k}\rangle\Bigr)+\cdots,
\label{eq:GS}
\ee
where the two N\'eel states ${\cal N}_A$ and ${\cal N}_B$ are the most probable states with $p_{\rm max}=|a_{\rm max}|^2$. The second most probables states having $p'_{\rm max}=|a'_{\rm max}|^2 <  p_{\rm max}$, labelled $|\varphi_{A/B,k}\rangle$ ($k \in [1,N_b]$), are obtained by applying one term $S_i^+S_j^- + S_i^-S_j^+$ on a nearest-neighbor bond $k=\langle i,j \rangle$ on one of the N\'eel states. They have two ferromagnetic domain walls (DW) which are as close as possible, {\it i.e.} at distance 2 (red spins) $|\cdots{\color{black}\uparrow\downarrow\uparrow \downarrow}{\color{red}\uparrow\uparrow}{\color{red}\downarrow\downarrow}{\color{black}\uparrow\downarrow\uparrow}\cdots\rangle$. We simply have $\sum_{k}|\varphi_{A/B,k}\rangle ={\cal H}_{od}|{\cal N}_{A/B}\rangle$. The dots in Eq.~\ref{eq:GS} represent
 states appearing with lower probabilities (either two DWs states with a larger separation between DWs, or states having more than two DWs), which are obtained by a higher number of application of ${\cal H}_{od}$. For the following calculation we can safely ignore such states.
The goal is to compute the first participation gap ${\cal G}$, defined as
\be
{\cal G}=\ln\left(\frac{p_{\rm max}}{p'_{\rm max}}\right).
\ee
To do so, we just need to compute $\langle {\cal N}_{A/B} | {\cal H} |\Psi_0\rangle= E_0 \langle {\cal N}_{A/B} | \Psi_0\rangle$, with $E_0$ the ground-state energy. The diagonal part ${\cal H}_d$ provides a ``classical" energy to each state in $|\Psi_0\rangle$. Upon applying ${\cal H}_{od}$ to $|\Psi_0\rangle$ (in the form of Eq.~\ref{eq:GS}), the two N\'eel states necessarily give $|\varphi_{A/B,k}\rangle$ states, whereas there are three possibilities for the $|\varphi_{A/B,k}\rangle$ states: either they create back ${\cal N}_{A/B}$, separate further the two DWS, or create two more DWs. Only the first case gives a contribution to $\langle {\cal N}_{A/B} | {\cal H} |\Psi_0\rangle$, and we readily obtain
\be
N_b\left(\frac{J_{xy}}{2}a'_{\rm max}-\frac{J_z}{4}a_{\rm max}\right)=E_0 a_{\rm max},
\ee
yielding
\be
{\cal G}=-\ln\left[ \left(\frac{2E_0}{J_{xy}N_b}+\frac{J_z}{2J_{xy}}\right)^2 \right].
\label{eq:exactG}
\ee
The classical energy of the N\'eel states being $E_{\rm cl}=-N_bJ_z/4$, $\cal G$ can be written as
\be
{\cal G}=-\ln\left[  \left(\frac{2(E_0-E_{\rm cl})}{J_{xy}N_b}\right)^2 \right].
\ee
From the above expression we find that the first participation gap $\cal G$ is finite for most systems,
and can be understood as a measure of how classical a system is. Indeed,  $\cal G \to \infty$ when the
ground-state energy $E_0\to E_{\rm cl}$, meaning that only the N\'eel configurations
$\ket{{\cal N}_{A/B}}$ have a finite weight in the ground-state. Conversely, when quantum
fluctuations are maximal, {\it e.g.} for a collection of $S=1/2$ dimers, we trivially get ${\cal
G}=0$. In Fig.~\ref{fig:Gaps}, we plot the numerical values of $\cal G$ for a few examples of  $S=1/2$ antiferromagnetic models, either having $SU(2)$ symmetry ($\Delta=1$ for the XXZ
Hamiltonian Eq.~\ref{eq:xxz}) as well as for the $S=1/2$ XXZ chain for various values of the easy axis anisotropy $\Delta$. From such numbers, it is not possible to conclude on the nature of the ground-states, and the only qualitative information which can be inferred is that $\cal G$ increases when antiferromagnetic (quasi)-order gets stronger.

For the XXZ chain, we have computed the participation gap ${\cal G}=\ln p_{\rm max}-\ln p'_{\rm max}$ directly using QMC simulations for $L=32$ sites with $\Delta\in[0,5]$ (blue crosses in Fig.~\ref{fig:Gaps}). Results perfectly match the analytical expression Eq.~\ref{eq:exactG} using the exact value for the GS energy $E_0$ from solving the Bethe Ansatz~\cite{Bethe31} equations on $L=32$ finite chains (red dashed line in Fig.~\ref{fig:Gaps}). Note also that a perturbative expression in the Ising limit $\Delta\gg 1$ can be obtained (see below Sec.~\ref{sec:xxzpert}): ${\cal G}\simeq 2\ln (2\Delta)$, which compares well (green line in Fig.~\ref{fig:Gaps}) with exact results for large enough anisotropies.

Given the relation between the PS and the ES of RK wave-functions (see Sec.~\ref{sec:basic}), we finally note an important consequence of the above argument: the ``entanglement gap" of RK wave-functions never closes (except in some decoupled limit), even if the system undergoes a quantum phase transition where the true gap vanishes. This has been observed numerically in the case of the square-triangular quantum dimer model~\cite{stephan_renyi_2012}, but we provide here a generic argument. This is in sharp contrast to several situations  (see {\it e.g.} Refs~\cite{calabrese_entanglement_2008,li_entanglement_2008,poilblanc_entanglement_2010, james_understanding_2013,Lauchli13}) where the ES is gapless. This failure of entanglement gap to mimic the behavior of the true gap for RK wave-functions has been recently used to question the universal content of the entanglement spectrum~\cite{chandran_how_2013}. 

\subsection{Full participation spectra from Quantum Monte Carlo}
\begin{figure}[h!]
\bc
\includegraphics[width=0.8\columnwidth]{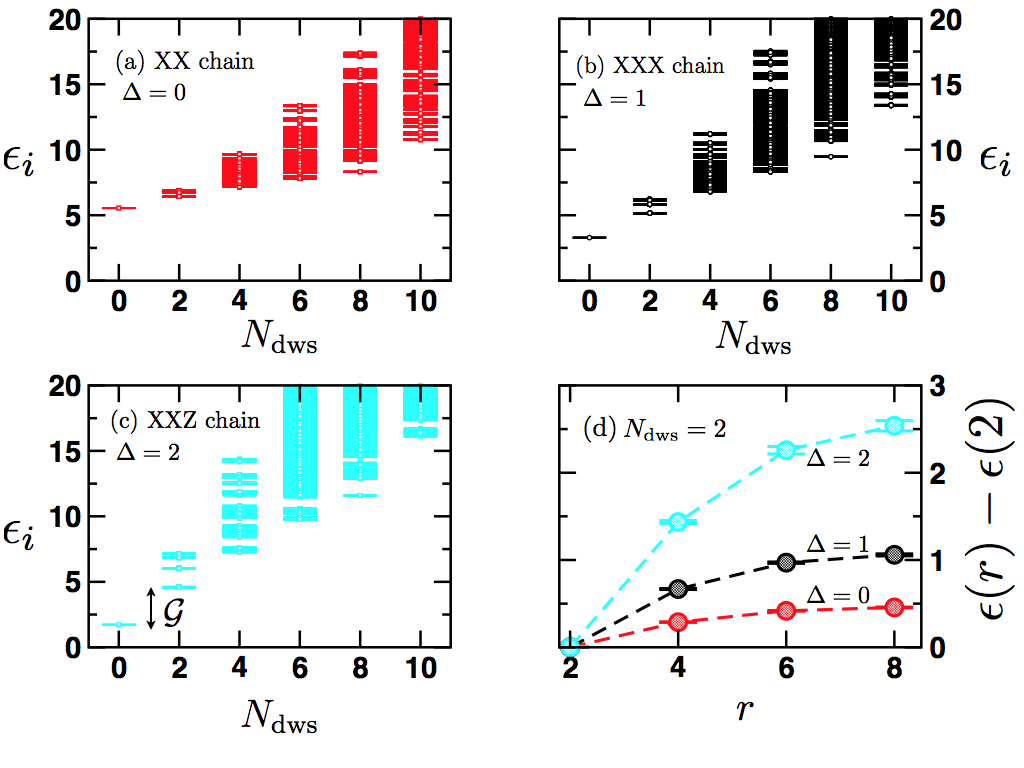}
\ec
\caption{Quantum Monte Carlo results for the participation spectra $\epsilon_i=-\ln p_i$ of various periodic spin-$\frac{1}{2}$ XXZ chains of $N=16$ sites for (a) $\Delta=0$, (b) $\Delta=1$, (c) $\Delta=2$, plotted against $N_{\rm dws}$ the number of ferromagnetic DWs. (d) Effective repulsion between 2 DWs for the three different anisotropies.}
\label{fig:occupation}
\end{figure}

As first discussed in Refs.~\cite{luitz_universal_2014,luitz_shannon_2014}, one can access the
pseudo-energies $\epsilon_i=-\ln p_i$ in a given computational basis using QMC simulations. We illustrate this with small XXZ chains ($L=16$ sites) for three representative points: free-fermion $\Delta=0$, Heisenberg $\Delta=1$, and Ising regime at $\Delta=2$, for which we display in Fig.~\ref{fig:occupation} the participation spectra recorded in the $\{S^z\}$ basis. For these particular cases, the GS belongs to the $S^z_{\rm tot}=0$ sector, the number of basis states is relatively small $16!/(8!)^2=12870$. 

The pseudo-energies $\epsilon_i$ can be displayed {\it vs.} their number of ferromagnetic domain walls
$N_{\rm dws}=0,~2,~4,\cdots, L-2$, as done in Fig.~\ref{fig:occupation} where we clearly see that
$\epsilon_i$ increases with $N_{\rm dws}$. The most probable states are the two N\'eel states with $N_{\rm dws}=0$, separated from $N_{\rm dws}=2$ states by the participation gap $\cal G$ computed above. The family of states having 2 DWs displays a substructure which can be understood in term of the relative separation between DWs, in a similar way as discussed in Ref.~\cite{luitz_shannon_2014}. As plotted in the panel (d) of Fig.~\ref{fig:occupation}, the pseudo-energies of the $N_{\rm dws}=2$ states increase with the distance $r$ between 2 DWs. A quantitative analysis of this effective repulsion is given below in Sec.~\ref{sec:DWS}.

For higher pseudo-energy states and larger numbers of DWs, an effective description based on pairwise DW interaction turns out to be much harder. For $N_{\rm dws}\ge 4$, the density of states increases, as well as the pseudo-energy packet width. Nevertheless a level repulsion is clearly visible when the Ising anisotropy increases as the size of the $N_{\rm dws}$-resolved packets gets larger with increasing $\Delta$.

\subsection{Effective pairwise interaction between domain walls}
\label{sec:DWS}
\begin{figure}[h!]
\bc
\includegraphics[width=0.62\columnwidth,clip]{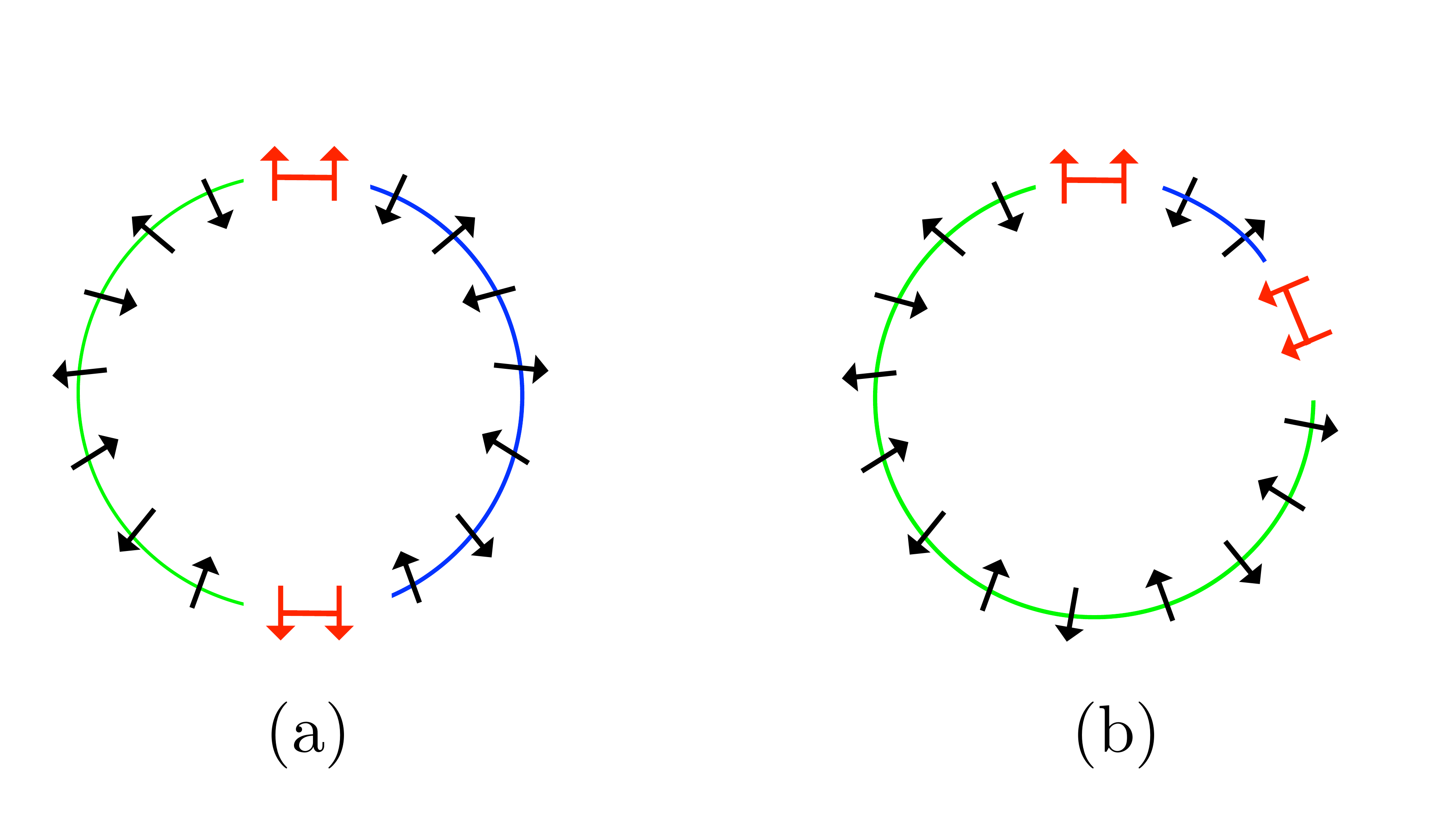}
\caption{Schematic picture for a period spin chain in a $\{S^z\}$ basis state having $N_{\rm dws}=2$. The two DWs (red) separate the two N\'eel patterns ${\cal N}_A$ and ${\cal N}_B$ (blue and green lines). If the separation $r$ is maximal (a) the staggered magnetization is zero whereas for smaller $r$ (b) the basis state yields a finite staggered magnetization.}
\label{fig:pict2dws}
\ec
\end{figure}

\begin{figure}[h]
\bc
\includegraphics[width=\columnwidth]{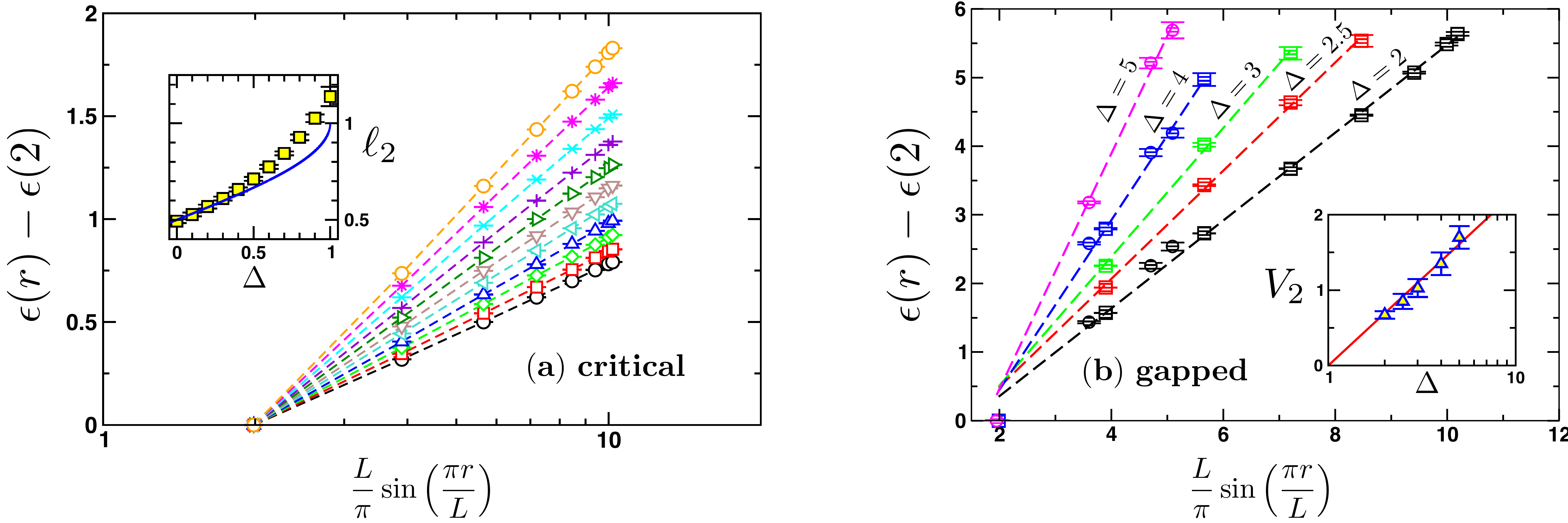}
\caption{Quantum Monte Carlo results for the effective pairwise repulsion $\epsilon(r)-\epsilon(2)$ between
    DWs obtained in the $N_{\rm dws}=2$ sector of the participation spectra for $L=32$ and $L=16$
    (circles in panel (b)) XXZ chains for various anisotropies $\Delta$ as indicated on the plots.
    (a) Logarithmic growth obtained over the full critical regime $\Delta\in [0,1]$, with a prefactor $\ell_2(\Delta)$ in Eq.~\ref{eq:l2} displayed in the inset {\it vs.} $\Delta$. The blue line is the prediction~\cite{JMPrivate} $\ell_2=1/(2K)$. (b) Linear confinement in the N\'eel ordered regime $\Delta> 1$. 
Inset: prefactor $V_2(\Delta)$ in Eq.~\ref{eq:V2} plotted {\it vs.} $\Delta$, the symbols are numerical estimates from linear fits in the main panel and the red line is the analytical perturbative estimate $V_2=\ln \Delta$.}
\label{fig:V2}
\ec
\end{figure}

At a qualitative level, the level repulsion observed in Fig.~\ref{fig:occupation}(d) can be simply explained following the sketch displayed in Fig.~\ref{fig:pict2dws}, where one sees that (a) when the distance between 2 DWs is maximal $r=L/2$, the two N\'eel patterns, ${\cal N}_A$ and ${\cal N}_B$ have the same size, resulting in a zero staggered magnetization for such a basis state. Conversely, when $r$ is small (b), one of the two N\'eel configurations prevails, which favour antiferromagnetic correlations. This phenomenology is expected to qualitatively discriminate GS having short-range  against long-range N\'eel order, as already discussed for line shaped subsystems in 2d antiferromagnets in Ref.~\cite{luitz_shannon_2014}. We also note that the charged particles Dyson-Gaudin gas representation developed in Ref.~\cite{stephan_shannon_2009} for the PS of $XXZ$ chains also agrees with this phenomenology.
In order to get more quantitative insights on the pairwise interaction between DWs we focus on $N_{\rm dws}=2$ and study the behavior of $\epsilon(r)-\epsilon(2)$, $r=2$ being the closest possible distance between 2 DWs, as a function of the Ising anisotropy for XXZ chains. QMC results for periodic chains of $L=32$ sites are shown in Fig.~\ref{fig:V2} for both critical quasi-long-range-ordered (QLRO) and gapped N\'eel ordered cases. In both regimes, the effective interaction is repulsive, as already understood from the simple above argument, but displays clearly distinct scalings as a function of the chord distance ${\tilde r}=L/\pi \sin(\pi r/L)$. Indeed, from our simulation we get the following scaling forms:
\bea
\label{eq:l2}
\epsilon(r)-\epsilon(2)&\sim&\ell_2\ln{\tilde r}~~~~{\rm (QLRO)}\\
&\sim& {V}_2 {\tilde r}~~~~~~~{\rm (NEEL)}
\label{eq:V2}
\eea
The linear confinement  $\sim V_2 {\tilde r}$ in the gapped regime can be understood  using perturbative arguments in the limit $1/\Delta\ll 1$, as we discuss below in Sec.~\ref{sec:xxzpert}. Indeed, when computing the corrections to the classical GS ${\ket{{\cal N}_{A/B}}}$, the separation of two DWs at a distance $r$ is found to be controlled by $(r/2)$-th order processes $\sim 1/\Delta^r$. As a result, $p_{\rm 2dws}(r)/p_{\rm max} \sim \Delta^{-r}$, yielding $\epsilon(r)\sim r\ln \Delta$. This linear confinement is nicely checked in Fig.~\ref{fig:V2} (b)  where $\epsilon(r)-\epsilon(2)$ displays the form Eq.~\ref{eq:V2} with $V_2(\Delta)$ remarkably well described by $V_2=\ln\Delta$ (red line in the inset of Fig.~\ref{fig:V2}(b)).

In the antiferromagnetic critical regime $\Delta\in[0,1]$, the absence of true long-range N\'eel order does not produce such a strong linear confinement for the DWs but a logarithmic confinement is rather observed in Fig.~\ref{fig:V2} (a) with a prefactor $\ell_2(\Delta)$ in Eq.~\ref{eq:l2} which is shown in the inset of Fig.~\ref{fig:V2} (a).  Using a scaling argument based on magnetic vertex operators in the free-field representation~\cite{JMPrivate}, $\ell_2$ is  expected to be directly related to the LL parameter by $\ell_2=1/(2K)$. In the inset of Fig.~\ref{fig:V2} (a) we verify this prediction already with relatively small system lengths ($L=32$). We have checked for representative values of $\Delta$ that deviations observed for $\Delta>0.5$ are due to finite-size effects~\cite{unpublished}.

\subsection{Perturbative results for the easy axis limit of the XXZ model}
\label{sec:xxzpert}
Several interesting features of the PS can be obtained perturbatively in the easy axis limit $\Delta\gg 1$ of the XXZ model Eq.~\ref{eq:xxz}, where the ground-state can be expanded as:
\be
|\Phi_0\rangle\propto |{{\cal{N}}_A}\rangle + |{{\cal{N}}_B}\rangle+\frac{1}{\Delta} \sum_n \alpha_n  |n\rangle+ \frac{1}{\Delta^2} \sum_{n'} \alpha_{n'}  |n'\rangle +\cdots,
\ee
where $|{\cal
N}_{A/B}\rangle$ the two N\'eel states and $|n\rangle, |n'\rangle$ are domain wall excitations above $|{\cal N}_{A/B} \rangle$ ($|n\rangle$ was previously denoted $|\varphi_{A/B,k}\rangle$ in Sec.~\ref{sec:gap}). As mentioned above, the perturbative processes which separate two DWs far apart, at a distance $r$, appear at order $r/2$, with a coefficient in the GS wave function $\sim \Delta^{-r}$. 

A straightforward calculation of the ground-state wave-function using second order perturbation theory provides the probability of the most probable states (the two classical N\'eel states) for a chain of $L$ spins:
\be
p_{\rm max}^{\rm 1d}=\frac{1/2}{{1+\frac{L}{4\Delta^2}+\frac{L^2}{32\Delta^4}+{\cal O}(\frac{L^3}{\Delta^{6}})}}.
\label{eq:pmax}
\ee
From such expansion one can conjecture the following exponential form:
\be
p_{\rm max}^{\rm 1d}=\frac{1}{2}\exp(-\frac{L}{4\Delta^2}), 
\ee
which implies for the $q=\infty$ R\'enyi entropy $S_\infty=\frac{L}{4\Delta^2}+\ln 2$. 

\begin{figure}[ht]
\bc
\includegraphics[width=0.55\columnwidth]{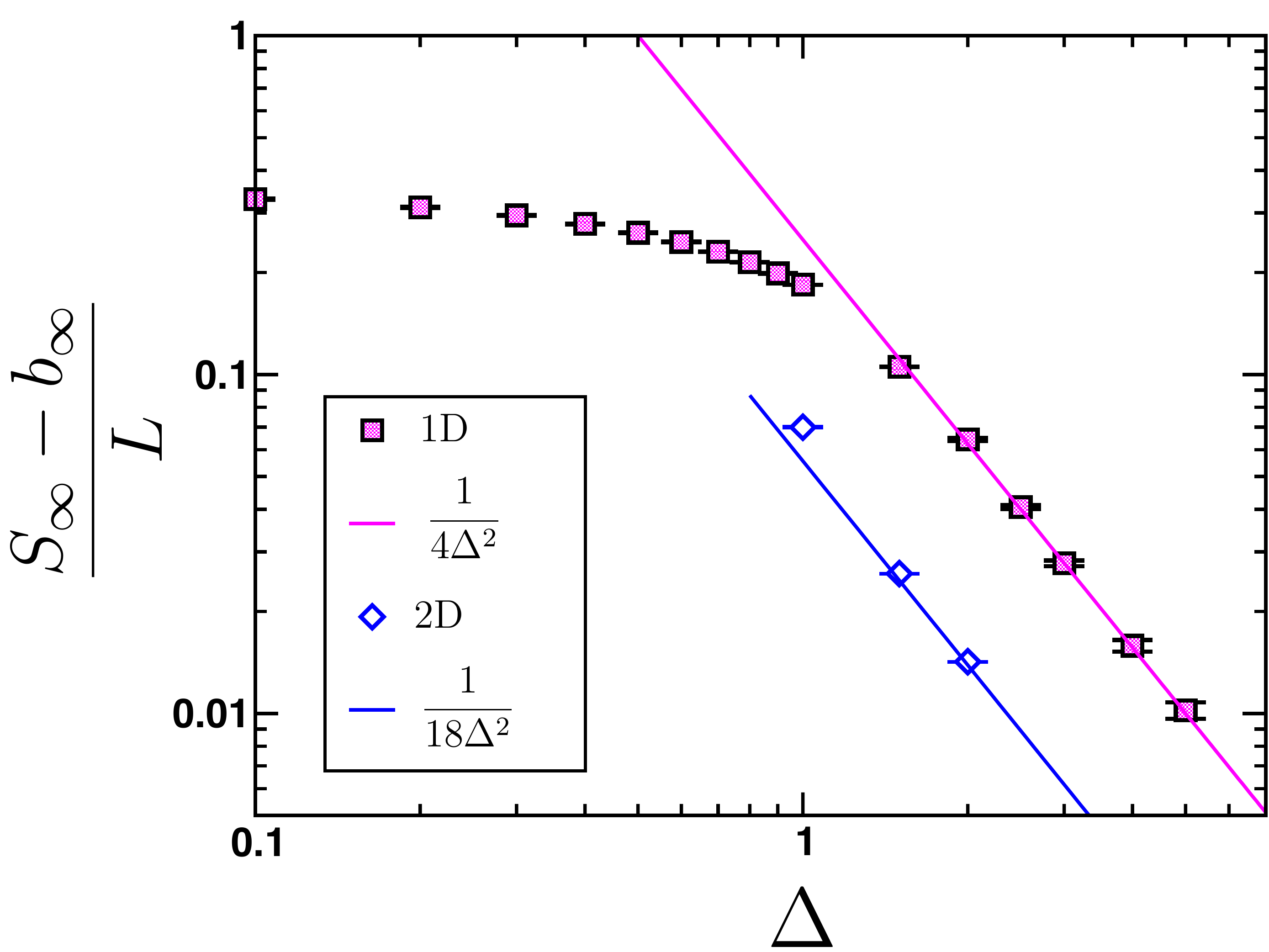}
\ec
\caption{QMC (symbols) and perturbative (lines) results for the leading coefficient $a_\infty=(S_\infty -b_{\infty})/L$ of the $q=\infty$ R\'enyi entropy of XXZ models in 1d and 2d (square lattice) plotted against the Ising anisotropy $\Delta$.}
\label{fig:ainfty}
\end{figure}

Already for $\Delta\ge 1.5$, this expression
is in very good agreement with QMC results as shown in Fig.~\ref{fig:ainfty} where $(S_\infty -b_{\infty})/L$ is plotted against $\Delta$ for $L=32$ XXZ chains with $b_\infty = \ln 2$. This result can be also extended to higher dimension, for instance for the 2d square lattice of $L$ sites we obtain 
\be
p_{\rm max}^{\rm 2d}=\frac{1}{2}\exp(-\frac{L}{18\Delta^2}),
\ee
which also agrees with QMC data obtained for a $20\times 20$ square lattice (blue symbols in Fig.~\ref{fig:ainfty}).

Finally, from such a perturbative expansion, one can discuss the participation gap $\cal G$ studied above in Sec.~\ref{sec:full}. From Eq.~\ref{eq:pmax}, it is straightforward to see that ${\cal G}\simeq 2\ln (2\Delta)$, as shown by the green line in Fig.~\ref{fig:Gaps}.
\section{Participation spectra for the line subsystem in a $d=2$ system}

\label{sec:sub}
\subsection{Participation spectra and entanglement Hamiltonians}
\subsubsection{General ideas}
We now turn to the study of participation spectra for a subsystem obtained by bipartition of the
total system. In particular, we will concentrate on $1d$ subsystems of $2d$ systems 
and discuss what the PS can bring on their ``entanglement Hamiltonians'' (see definition below). 
Accessing all the entries of the reduced density matrix is notoriously hard using QMC for spin- or
bosonic systems (even for systems with no sign problem), as recently discussed by Chung
and co-workers~\cite{chung_entanglement_2013} who tried to reconstruct the entanglement spectrum
using QMC estimates of the trace of the first $n^{\rm th}$ powers of the reduced density matrix~\cite{Song12}. This task is extremely difficult to achieve and practically restricted to the very bottom of the entanglement
spectrum. In this context, the relevance of the low-lying part of the entanglement spectrum to understand the actual
physical properties of the full system has been questioned~\cite{chandran_how_2013}.  In particular,
the effective inverse temperature $\beta_{\rm eff}$ which appears in the definition of the entanglement Hamiltonian ${\hat{\cal H}_E}$
\be
{\hat{\rho}}_B=\frac{\exp(-\beta_{\rm eff}{\hat{\cal H}}_{E})}{Z},
\label{eq:HE}
\ee
where $\hat \rho_B=\Tr_A \ket{\Psi_0} \bra{\Psi_0}$ is the reduced density matrix of the subsystem $B$, is a relevant quantity regarding the nature of the quantum state of matter. Note that in the above definition, the energy scale of the entanglement Hamiltonian is set to $J=1$ ({\it i.e.} the effective temperature $T_{\rm eff}=\beta_{\rm eff}^{-1}$ is given in units of $J$).
 \begin{figure}[b]
     \bc
     \includegraphics[width=.6\columnwidth]{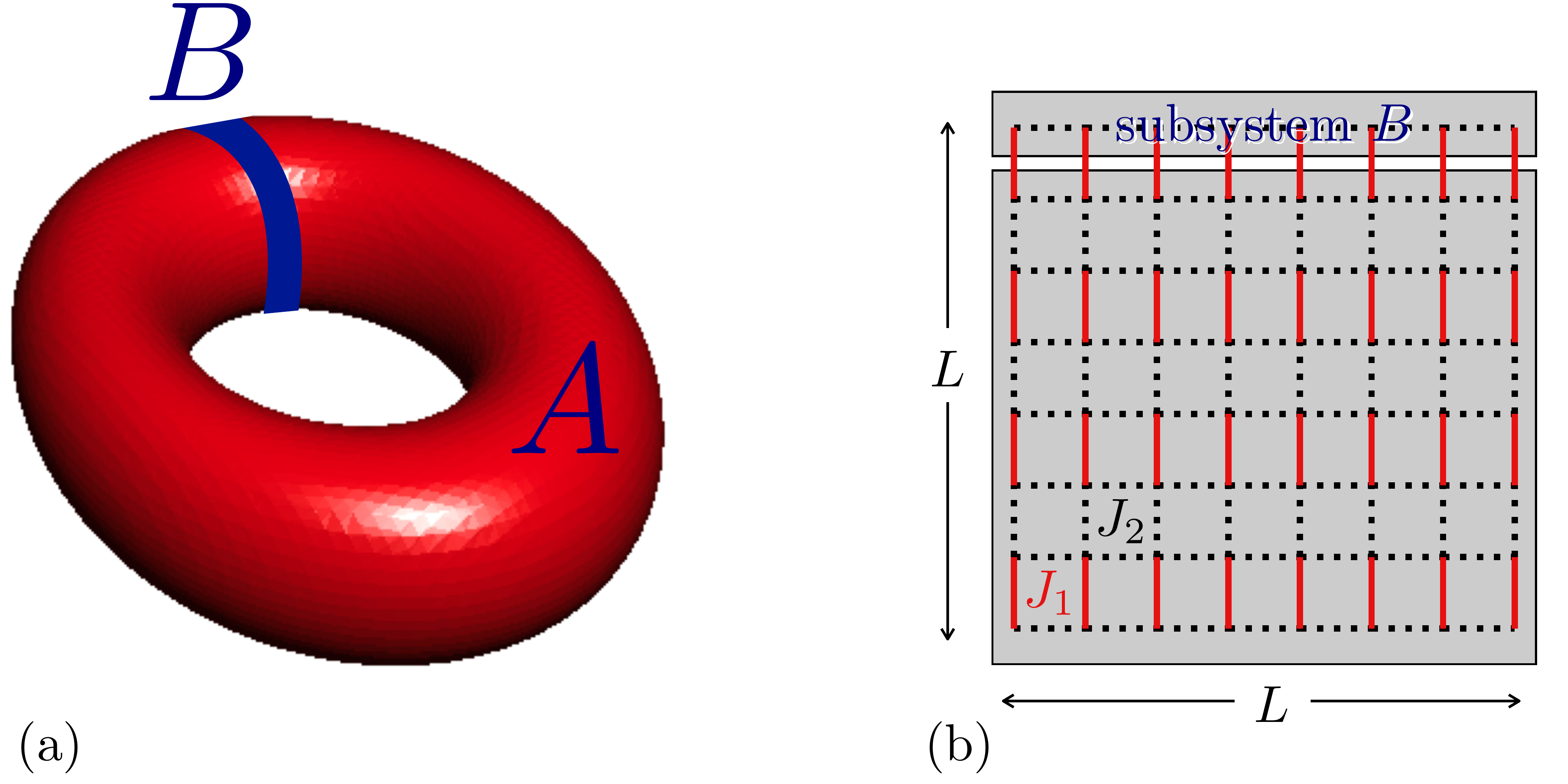}
     \caption{Schematic picture for the chosen line-shaped bipartition. (a) Subsystem $B$ is a
     single line of length $L$ embedded in a $L\times L$ torus. (b) The dimerized Heisenberg lattice
 model Eq.~\ref{eq:Hdim} has thick (red) lines for strong bonds with coupling $J_1$ and weak interdimer couplings $J_2\le J_1$ represented by dotted (black) lines.}
     \label{fig:torus}
     \ec
 \end{figure}

We want to address the question of the effective Hamiltonian in a subsystem $B$ consisting of a one dimensional line of spins (as depicted in Fig.~\ref{fig:torus}), not by computing the spectrum of the reduced density matrix $\hat \rho_B$ (due to the difficulties mentioned above), but by focusing on the PS of $\hat \rho_B$ in the $\{\ket i\}=\{ S^z\}$ basis. The PS of $B$ is defined using the diagonal elements of the reduced density matrix
\be
\epsilon_i^{B}=-\ln\left(\langle i|{\hat{\rho}}_B|i\rangle\right).
\ee
Therefore, using the entanglement Hamiltonian definition of Eq.~\ref{eq:HE}, one can define an effective PS
\be
\epsilon_i^{E}=\ln Z-\ln\left(\langle i|\exp(-\beta_{\rm eff} {\hat{\cal H}}_{E}) |i\rangle\right),
\ee
which has to fulfil for all levels $i$
\be
\epsilon_i^E=\epsilon_i^B,
\label{eq:fulfil}
\ee
if ${\cal H}_{E}$ is indeed the correct entanglement Hamiltonian and $T_{\rm eff}=\beta_{\rm eff}^{-1}$ the effective temperature.
\subsubsection{2d dimerized model}
In practice we focus on a 2d quantum spin-$\frac{1}{2}$ dimerized Heisenberg model  defined on a $L\times L$ square lattice (see Fig.~\ref{fig:torus}(b)) by the Hamiltonian
\begin{equation} 
H_{\rm dimer}= J_1 \sum_{{\rm dimers}} \vec{S}_i \cdot
\vec{S}_j +J_2 \sum_{\rm{links}} \vec{S}_i \cdot \vec{S}_j, 
\label{eq:Hdim}
\end{equation}
with $J_1,J_2\geq 0$ and where the two terms correspond to the summation over stronger bonds for columnar dimers and to the summation over the weaker links between these entities. We will only consider
$g=J_2/J_1\leq1$ here, with $g=1$ yielding the isotropic Heisenberg antiferromagnet on the square
lattice. This model has been intensively studied at zero
temperature~\cite{troyer_97,matsumoto_ground-state_2001,wang_high-precision_2006,albuquerque_quantum_2008,wenzel_09,sandvik_computational_2010} and exhibits a $2d+1$ $O(3)$ quantum critical point at $g_c=0.52370(1)$\cite{sandvik_computational_2010} separating a disordered gapped phase for $g<g_c$  from an antiferromagnetic N\'eel long-range ordered phase which occurs at $g>g_c$, with a spontaneous breaking of $SU(2)$ symmetry. 

We have already discussed SR entropies and PS for such a line shaped subsystem in Ref.~\cite{luitz_shannon_2014} where several results have been obtained for the universal scaling properties of $S_\infty^{\rm line}$ across the phase diagram $0\le g \le 1$ of model Eq. \ref{eq:Hdim}. Concerning the PS, ordered and disordered phases are qualitatively different, in particular regarding the effective interaction between ferromagnetic DWs (see also Sec.~\ref{sec:DWS}). Concretely, these objects experience a pairwise repulsive interaction which grows linearly with the distance (linear confinement) in the N\'eel regime while the confinement is much weaker, and short-ranged, in the gapped phase, with a deconfinement occurring above a finite distance controlled by the finite correlation length of the disordered phase.

In the following we want to address the question of which effective entanglement Hamitonian  correctly describes the entire PS, trying to satisfy Eq.~\ref{eq:fulfil}, across the phase diagram of the dimerized antiferromagnetic model Eq.~\ref{eq:Hdim}. Our approach will consist in trying to compare different possible entanglement Hamiltonians, which are motivated by symmetry, perturbative arguments in the limit of small $g$, and also by the fact that they should display antiferromagnetic ordering at the effective temperature $T_{\rm eff}$ for $g>g_c$.
\subsubsection{Quantitative approach to compare two spectra: R\'enyi and Kullback-Leibler divergences} 
\label{sec:KL} 

For a quantitative comparison between the PS $\{\epsilon_i^B\}$ of subsystem $B$
 and  $\{\epsilon_i^E\}$ of the effective entanglement Hamiltonian, 
it is necessary to introduce a measure of distance between two such PS. 
This question is also strongly relevant for the ES, for which  the method presented here is also directly applicable.
The comparison of two PS translates mathematically to the problem of comparing two (in this case discrete) probability distributions $P$ and $Q$.
 R\'enyi introduced the \emph{R\'enyi divergence} of order $q$ 
 \begin{equation}
     I_q(Q|P) = \frac{1}{1-q} \ln \left( \sum_{i} \frac{Q_i^q}{P_i^{q-1}} \right),
 \end{equation}
 a quantity representing \emph{``the information of order $q$ obtained if the distribution $P$ is
 replaced by the distribution $Q$''}\cite{renyi1961}. Clearly, $I_q(Q|P)$ vanishes if the two
 distributions are identical. As for the case of the R\'enyi entropies, the R\'enyi divergence
 reduces to the classical result by Kullback and Leibler (KL)\cite{kullback1951} in the limit of $q\to 1$:
 \begin{equation}
     I_1(Q|P) = \sum_i Q_i \ln \frac{Q_i}{P_i}.
     \label{eq:I1}
 \end{equation}

Let us emphasize the quantitative information brought by KL and R\'enyi divergences which compare the two spectra {\it state by state} (including possible multiplicities), which contrasts with the qualitative information gained by a visual
comparison of spectra. In the following analysis, we will display mostly results for $I_1(\{\epsilon^B\}|\{\epsilon^E\})$ which
allows to compare PS across their entire range, as opposed to $I_{q\gg 1}$ which increases the
weight in the low ``pseudo-energy'' part (corresponding to higher probabilities). 
However, it is important to emphasize that we always check that the analysis remains stable under
variations of $q$.

For a reliable comparison of R\'enyi divergences for different model parameters and in particular
system sizes, we find that it is useful to consider the \emph{relative} R\'enyi divergences 
\begin{equation}
    \frac{I_q(Q|P)}{S_q(Q)}.
\end{equation}
This quantity denotes the relative information gain if the distribution $P$ is replaced by $Q$ with
respect to the information contained in $Q$.
\subsection{Effective entanglement Hamiltonian for the gapped regime}

  We now study the PS of the line shaped subsystem $B$ of the Hamiltonian given by Eq. \ref{eq:Hdim} in 2d and
  compare it to PS of 1d models which may be understood as effective entanglement Hamiltonians. All PS are obtained using the QMC technique introduced in Ref.~\cite{luitz_universal_2014,luitz_shannon_2014}.

\subsubsection{Short-ranged models}

Deep in the gapped phase, $J_2/J_1\ll 1$ one can apply perturbation theory to extract the
entanglement Hamiltonian ${\cal H}_E$ defined on the 1d subsystem $B$ (Fig.~\ref{fig:torus}),
similarly to what was done in previous works~\cite{Peschel11,lauchli_entanglement_2012,Chen13,Lundgren13} for 2-leg ladders~\cite{poilblanc_entanglement_2010}. At first order,  the
calculation for the 2d dimerized model is identical to the ladder
case~\cite{lauchli_entanglement_2012}, yielding the simple result (with fixed $J_1=1$):
\be
{\cal H}_E=\sum_{i\in B}\vec{S}_i \cdot
\vec{S}_{i+1},~~~~{\rm and}~~~~T_{\rm eff}=\frac{1}{2J_2}.
\label{eq:naive}
\ee
This is simply a $S=\frac{1}{2}$ Heisenberg chain problem at finite temperature $T_{\rm eff}$ which is well known to harbor short-range correlations with a finite correlation length $\sim 1/T_{\rm eff}$, in perfect agreement with the bulk correlation length of the dimerized model which grows linearly with $J_2$, deep in the gapped regime.
As in the ladder case, when $J_2$ increases (the bulk gap decreases) we expect non-negligible longer
range interactions in ${\cal H}_E$, as well as multi-spin processes~\cite{Cirac12} to arise from
higher order perturbation theory. Restricting our
study to two-body effects, we consider the following ``$\xi$-model'' as a potential entanglement Hamiltonian:
\be
{\cal H}_E(\xi)= -\sum_{\stackrel{i,j\in B}{i>j} }(-1)^{r_{ij}}{\rm
e}^{-\frac{(r_{ij}-1)}{\xi_E}}\vec{S}_i \cdot \vec{S}_{j}, 
\label{eq:xi}
\ee
where $r_{ij}$ is the minimal distance between sites $i$ and $j$: $r_{ij}={\rm
min}(|i-j|,L-|i-j|)$. For $\xi_E= 0$, we simply consider the Heisenberg chain given by Eq.~(\ref{eq:naive}).
 The $\xi$-model is non-frustrated and belongs to the class of short-range
models since it displays identical low-energy physics and is described by the same field
theory~\cite{giamarchi_quantum_2004} as the nearest-neighbor Heisenberg spin chain. Nevertheless, non-universal details are expected to depend on $\xi_E$, in particular the PS. 
\subsubsection{First example in the gapped phase}

  \begin{figure}[b]
  \centering
     \includegraphics[width=\columnwidth]{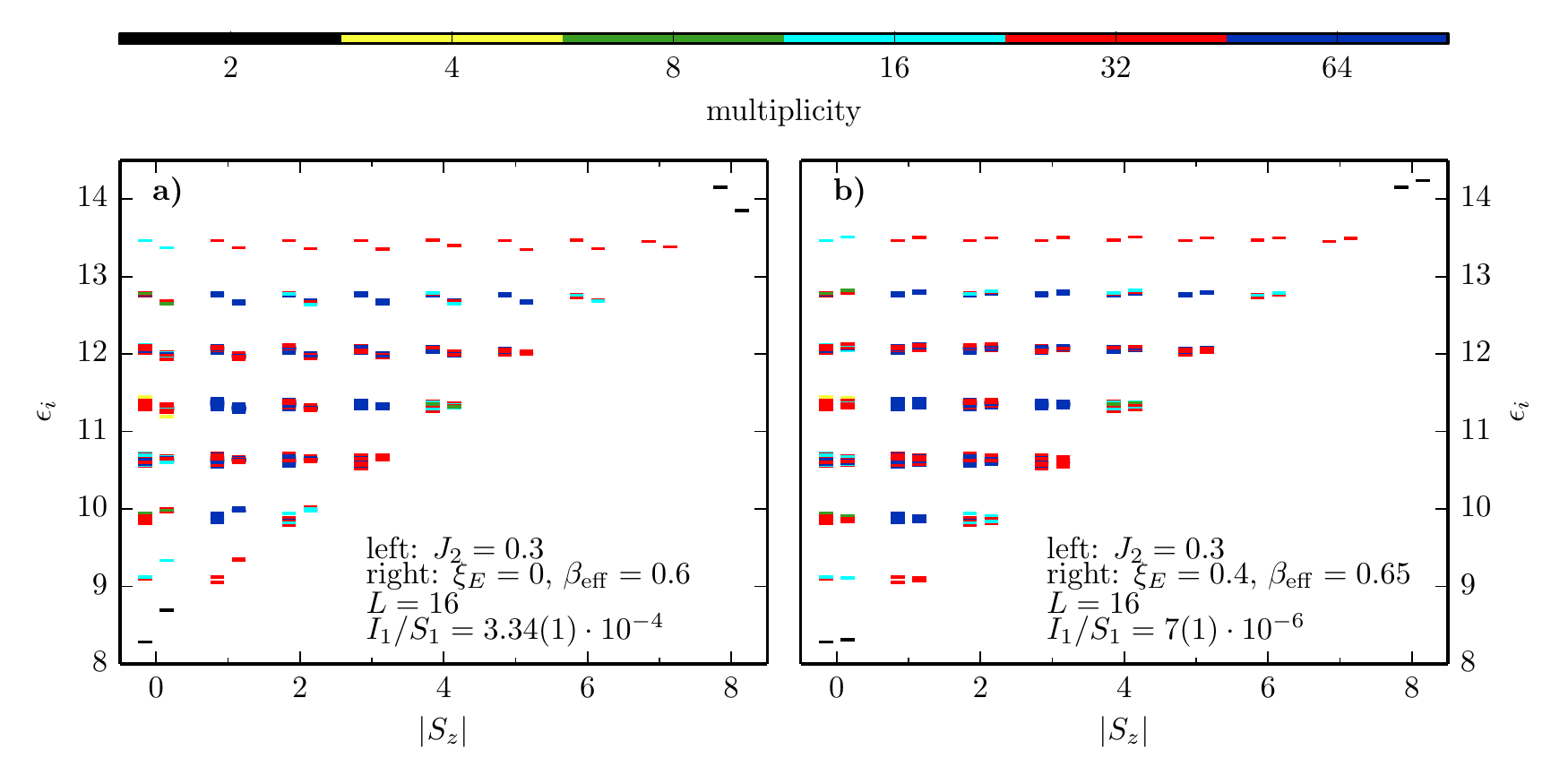}
     \caption{Comparison of participation spectra of the line subsystem for $J_2=0.3$ and $L=16$
         with the effective models. For each $|S_z|$ sector, two spectra are displayed (the left one corresponds to the line subsystem, the right to the effective model). On the left panel a), the effective model is the nearest-neighbor only spin chain model ($\xi_E=0$) at $\beta_{\mathrm{eff}}=0.6$. On the right panel b), it is the improved effective model with $\xi_E=0.4$
 and $\beta_{\mathrm{eff}}=0.65$. Errorbars are smaller than the linewidth and the different colors code for the different multiplicities of the basis states.}
     \label{fig:spectra_gap}
 \end{figure}

We first discuss the example of the dimerized 2d model at $J_2=0.3$ for which the PS is displayed in
Fig.~\ref{fig:spectra_gap}. For clarity, we display the spectrum for each $S^z$ sector for a line of
$L=16$ sites embedded in a $16\times 16$ lattice. Following the first order perturbation
result Eq.~\ref{eq:naive} we have superimposed the effective PS $\{\epsilon^E\}$ of a Heisenberg chain
with $\beta_{\rm eff}=2J_2=0.6$ to the actual PS $\{\epsilon^B\}$ of subsystem B in panel (a) of
Fig.~\ref{fig:spectra_gap}. The visual comparison is correct, respecting the multiplicities of the
levels, but the agreement can be significantly improved if a finite range $\xi_E>0$ is allowed using
the $\xi$-model Eq.~\ref{eq:xi}. The right panel (b) of Fig.~\ref{fig:spectra_gap} shows the
``best'' effective PS (in the sense of giving rise to the smaller R\'enyi divergences $I_q$), obtained with $\xi_E=0.4$ and $\beta_{\rm eff}=0.65$. 

\subsubsection{Evolution of the entanglement Hamiltonian across the full gapped regime}

 \begin{figure}[b]
     \centering
     \includegraphics[width=\columnwidth]{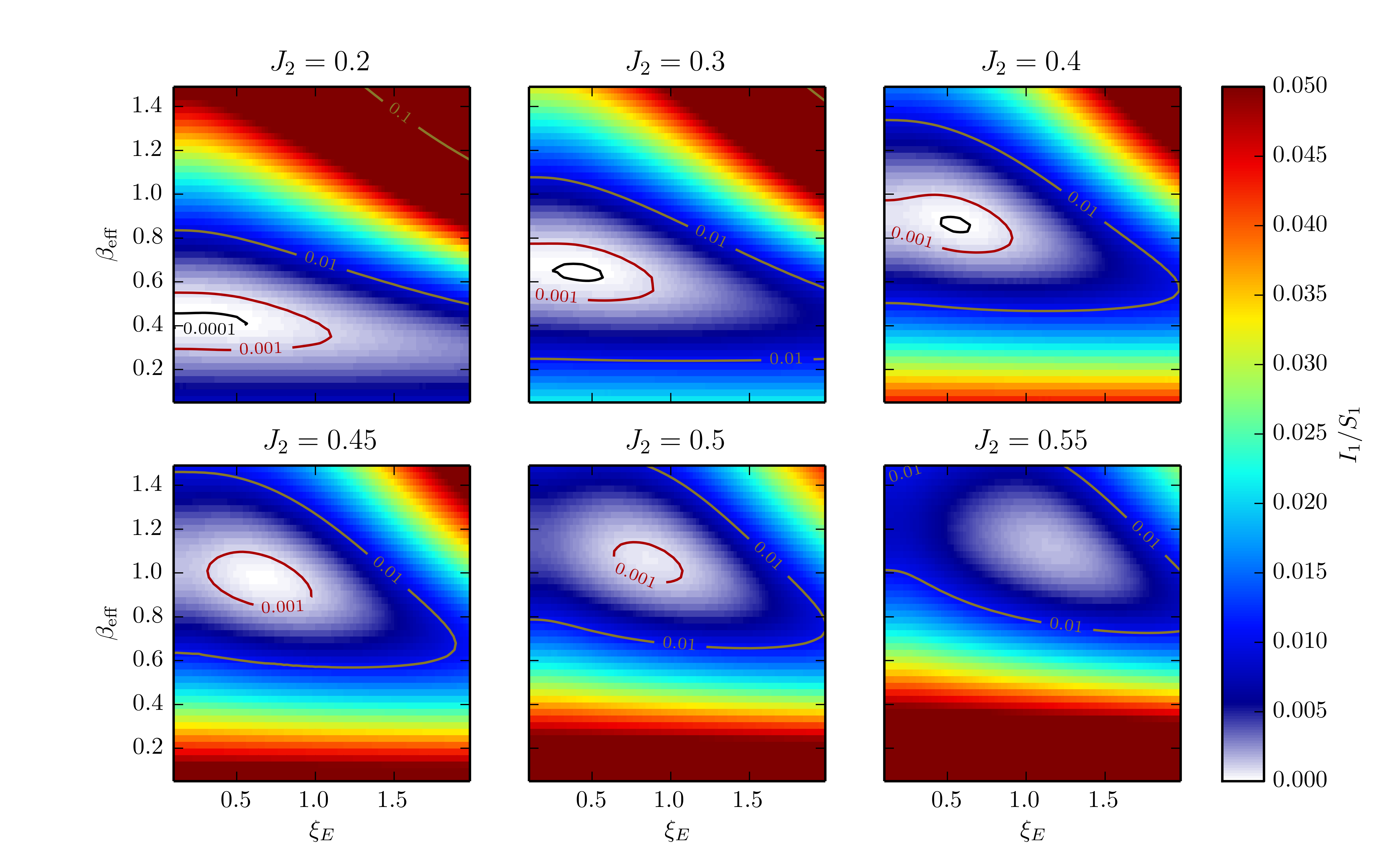}
     \caption{Relative KL divergence $I_1/S_1$ of the line shaped subsystem spectrum and
     the effective short range Hamiltonian ($\xi$-model) for different inverse temperatures $\beta_{\rm eff}$
     and ranges $\xi_E$ of the
     interaction. Here $L=16$. As $J_2$ approaches the phase transition, the range of the effective interaction and the inverse effective temperature both slowly increase. 
 }
     \label{fig:min_div_xi}
 \end{figure}

In order to monitor the evolution of the effective parameters of the entanglement Hamiltonian for
the entire disordered phase $0\le J_2 <J_c=0.5237$ we have scanned the two-dimensional parameter space
$\xi_E$ --- $\beta_{\rm eff}$, where a very large number of PS have been recorded using QMC
simulations of the $\xi$-model (Eq.~\ref{eq:xi}) for $L=16$ chains. Such effective PS are then
compared to the actual $\{\epsilon^B\}_{J_2}$ with $L=16$ for various values of $J_2$, as displayed
in Fig.~\ref{fig:min_div_xi} where color maps of the normalized KL divergences $I_1/S_1$ are shown 
(here $S_1$ is the Shannon entropy of the line $S_1=\sum_i \epsilon_i^B\exp(-\epsilon_i^B)$ in the
dimerized model). Before the calculation of the KL divergence, we use all translation symmetries of
the lattice in order to improve the quality of the spectra. A detailed bootstrap analysis reveals
that the error bars of relative KL divergences $I_1/S_1$ are typically smaller than $10^{-6}$ and
can be neglected in this discussion.
We clearly see in Fig. \ref{fig:min_div_xi} that a small area develops in the diagrams where
$I_1/S_1$ is extremely small, with a relative KL divergence between two spectra as small as
$I_1/S_1=0.001\%$. The parameter region with minimal KL divergence where both PS
$\{\epsilon^B\}_{J_2}$ and $\{\epsilon^E\}_{\xi_E,\beta_{\rm eff}}$ are almost identical, slowly
moves towards the upper right corner of the parameter space when $J_2$ increases while at the same
time, its size gradually shrinks to zero. For instance, in the last panel (bottom right) of
Fig.~\ref{fig:min_div_xi} for which the inter-dimer coupling is beyond the critical point
$J_c=0.5237$, the very bright region has disappeared (\textit{cf.} isodivergence lines) which signals that the $\xi$-model is not anymore appropriate as an entanglement Hamiltonian.

The fact that $\xi_E$ and $\beta_{\rm eff}$ both increase with $J_2$ can be qualitatively understood
as follows. When $J_2$ increases in the dimerized system, antiferromagnetic correlations build up
over an increasing range and in the same time, the finite energy gap decreases. We therefore expect
the coupling range in ${\cal H}_E$ to increase ($\xi_E$ grows) and the effective temperature
$1/\beta_{\rm eff}$ to decrease when $J_2$ increases.  More quantitatively, the first order
perturbative result (Eq.~\ref{eq:naive}) gives $\beta_{\rm eff}=2J_2$, as nicely checked in
Fig.~\ref{fig:min_div_comp}(b) where the optimal effective inverse temperature is plotted against $J_2$. Interestingly, we see that the first order perturbative result gives a very good description in the full gapped regime, and remarkably, the effective temperature remains finite when $J_c$ is approached. We will return to this in Sec.~\ref{sec:discussion}.

\begin{figure}[h]
     \centering
     \includegraphics[width=.65\columnwidth]{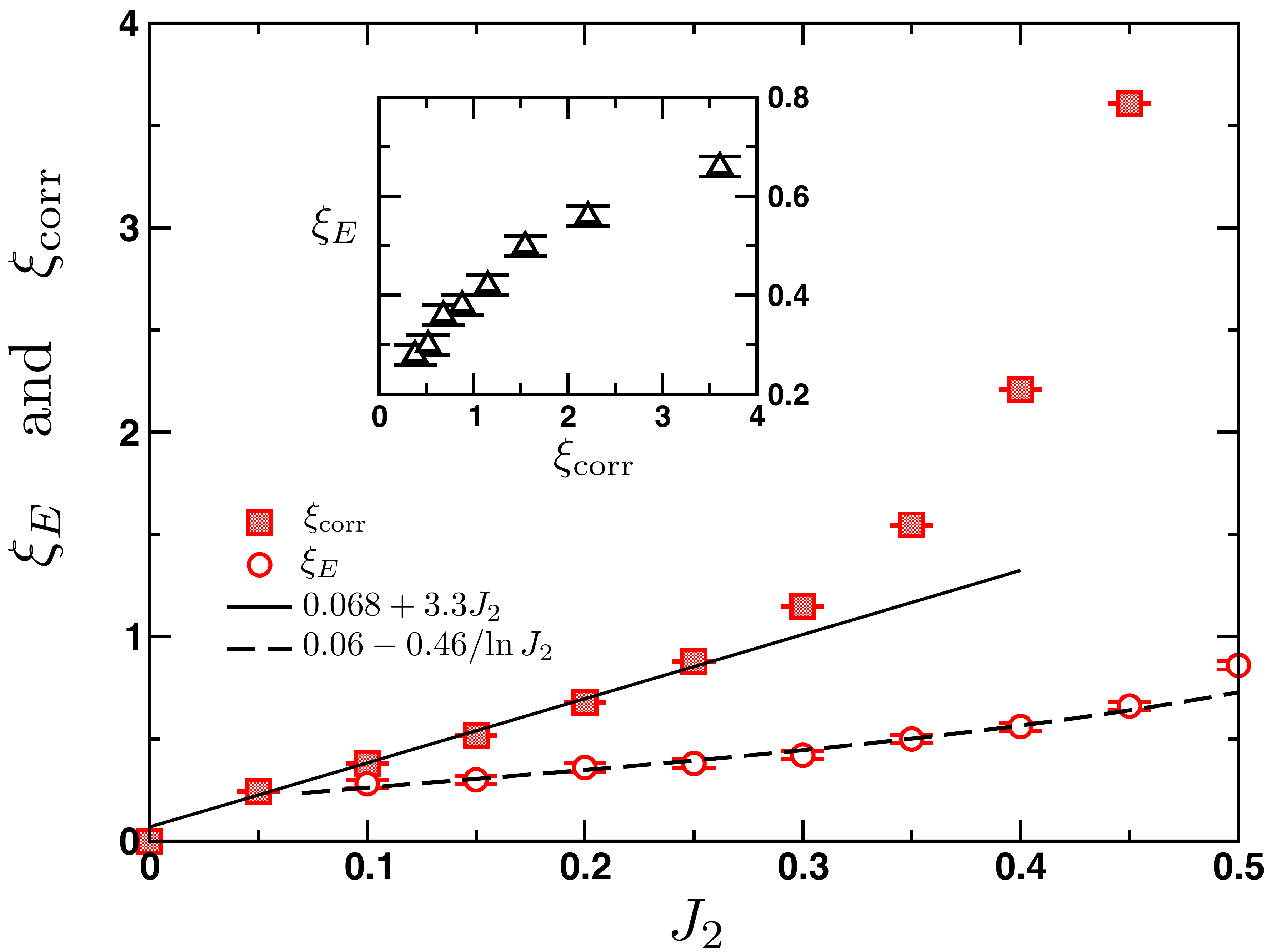}
     \caption{Evolution of both entanglement $\xi_E$ (from $L=16$ simulations) and correlation $\xi_{\rm corr}$ lengths (see text for definitions) against the inter-dimer coupling $J_2$. The inset shows using the same data the dependence of $\xi_E$ versus $\xi_{\rm corr}$.}
     \label{fig:xis}
 \end{figure}

The exponential form of the interactions in the $\xi$-model Eq.~\ref{eq:xi} can be simply understood following Ref.~\cite{lauchli_entanglement_2012}. Indeed, couplings at distance $r>1$ are generated at $r$-th order in perturbation theory, and are proportional to 
$(J_2)^r=\exp(-r|\ln J_2|)$ if $J_2\ll 1$. Therefore, in the small $J_2$ limit we expect the following behavior for the entanglement length
\be
\xi_E\propto -\frac{1}{\ln J_2}.
\ee
This has to be contrasted with the ``true'' correlation length $\xi_{\rm corr}$ of the gapped model which in the small $J_2$ limit grows linearly.  
In Fig.~\ref{fig:xis}, both the entanglement $\xi_E$ and the correlation length $\xi_{\rm corr}$ (measured using a second moment method~\cite{cooper_solving_1982}) are plotted  {\it{vs.}} $J_2$ in the gapped regime. At small $J_2$, $\xi_{\rm corr}$ shows a linear behavior and  $\xi_E$ is well described by the above expression $\sim |\ln J_2|^{-1}$, not only at small $J_2$ but also quite close to the critical regime. On the other hand, when the critical point is approached, $\xi_{\rm corr}$ clearly diverges much faster than $\xi_E$ which remains of order 1. The inset of Fig.~\ref{fig:xis} shows the mutual dependence of these two correlation lengths, suggesting a possible log dependence of $\xi_E$ on $\xi_{\rm corr}$ at large $J_2$. With the lattice sizes at hand, it is however hard to provide an estimate of how $\xi_E$ will behave when reaching the quantum critical point. Nevertheless we expect $\xi_E$ to diverge at the critical point where spin correlations are algebraic.

\subsection{Entanglement Hamiltonian in the N\'eel ordered phase}
In the N\'eel phase, the situation is more complex. Indeed, the effective 1d model for subsystem $B$
has to break $SU(2)$ symmetry at an effective temperature $T_{\rm eff}$. Therefore ${\cal H}_E$ has
necessarily to be long-ranged such that, as emphasized in Refs.~\cite{Alba12,kolley_entanglement_2013}, the low-energy part of the entanglement spectrum for continuous symmetry broken phases displays a tower of state (TOS) structure~\cite{Bernu92}. 

\subsubsection{$\alpha$-model}

The simplest Hamiltonian which exhibits such a TOS structure is the Lieb-Mattis
model~\cite{lieb_ordering_1962}. More generally, we will focus on the following non-frustrated power-law decaying spin chain Hamiltonian
\be
{\cal H}_E(\alpha)=\sum_{i\in B}\sum_{j\in B}\frac{(-1)^{r_{ij}}}{r_{ij}^{\alpha}}\vec{S}_i \cdot
\vec{S}_{j},
\label{eq:alpha}
\ee
which we denote as the $\alpha$-model. This Hamiltonian, intensively studied in Refs.~\cite{yusuf_spin_2004,laflorencie_critical_2005,beach_fractal_2007}, displays a rich phase
diagram, with a (zero temperature) quantum phase transition at $\alpha_c\simeq 2.2$~\cite{laflorencie_critical_2005}
between a N\'eel ordered phase for $\alpha< \alpha_c$ and a QLRO phase for $\alpha>\alpha_c$. Note
that for $\alpha=0$ the Lieb-Mattis model is recovered (the Lieb-Mattis model is usually defined with a prefactor $1/L$ to ensure energy extensivity, see below).

\begin{figure}[h]
     \centering
     \includegraphics[width=\columnwidth]{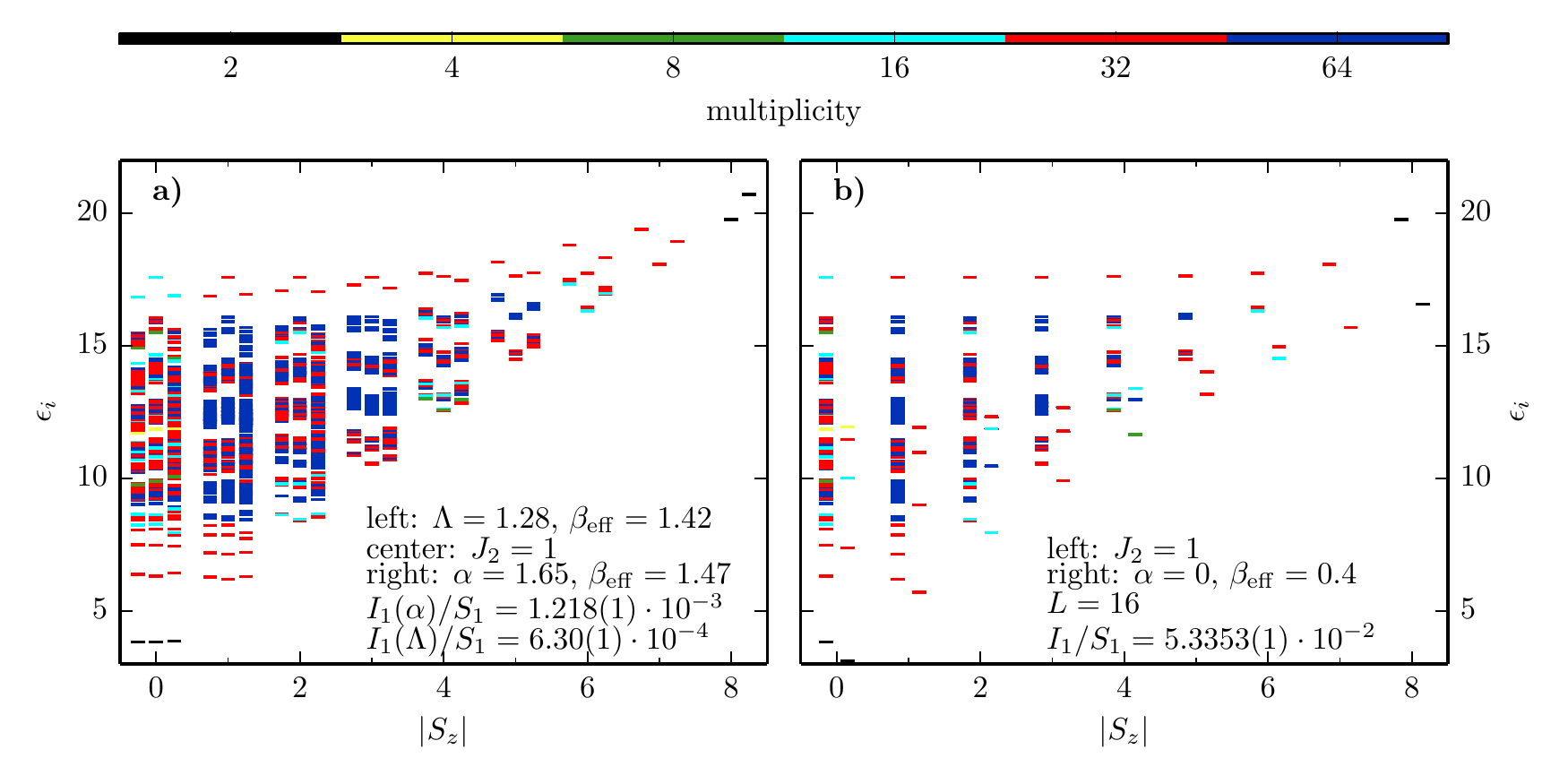}
     \caption{Comparison of participation spectra of the line subsystem for $J_2=1$ and $L=16$ with various
         effective models. Errorbars are smaller than 
 the linewidth and the different colors show the different multiplicities of the basis states. Right: comparison with the Lieb-Mattis model  at $\beta_{\rm eff}=0.4$. Left: Double comparison with (i) the $\alpha$-model (right $\alpha=1.65,~\beta_{\rm eff}=1.47$)  and (ii) the $\Lambda$-model (left $\Lambda=1.28,~\beta_{\rm eff}=1.42$).{}
 }
     \label{fig:spectra_lro}
 \end{figure}

As previously done (above in Fig.~\ref{fig:spectra_gap}) for the gapped regime, we now compare
the entire PS for the ordered phase at the isotropic point $J_2=1$ with PS of various effective
entanglement Hamiltonians. First, in the right panel of Fig.~\ref{fig:spectra_lro} we clearly see
that the LM model ($\alpha=0$), even for the lowest KL divergence found for $\beta_{\rm
eff}=0.4$, does not compare well with the spectrum of the line shaped subsystem $B$. The
comparison is clearly off at both quantitative ($I_1/S_1=5.3353(1)\cdot 10^{-2}$) and qualitative levels since the spectral  structures are very distinct.
On the other hand, the comparison in the left panel of  Fig.~\ref{fig:spectra_lro} gives much better accordance. There we see first on the right columns of the packets that the $\alpha$-model yields a very nice agreement using the following optimal parameters $\alpha=1.65$ and $\beta_{\rm eff}=1.47$. We have checked that in the thermodynamic limit the 1d power-law model at $\alpha=1.65$ remains N\'eel ordered at small temperature below $T_c\simeq 1/1.4$, which ensures that the line sub-system is ordered, as it should be. 

\begin{figure}[h]
     \centering
     \includegraphics[width=\columnwidth]{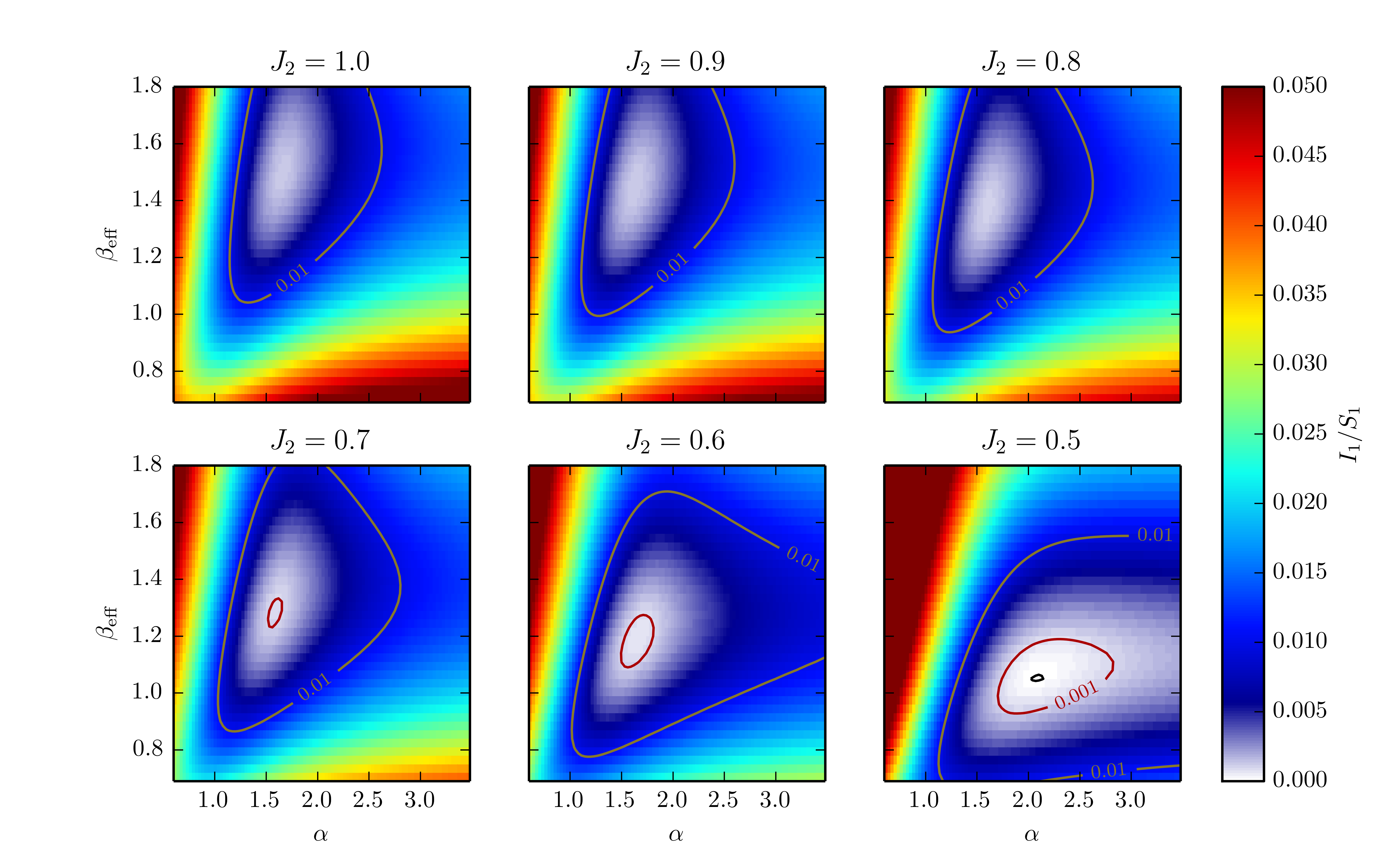}
     \caption{ Relative KL divergence $I_1/S_1$ of the line shaped subsystem spectrum and
     the effective long-range Hamiltonian ($\alpha$-model) for different inverse temperatures $\beta_{\rm eff}$
     and ranges $\alpha$ of the
     interaction. Here $L=16$. In the N\'eel phase, the optimal inverse temperature decreases with $J_2$ while the optimal $\alpha$ remains $\sim 1.6$. When one has crossed the phase transition (bottom right panel with $J_2=0.5$), the minimum of the range of the interaction $\alpha$ has abruptly changed to $\sim 2$.}
     \label{fig:min_div_alpha}
 \end{figure}

As done previously for the disordered case, we can follow the evolution of the effective parameters
of the entanglement Hamiltonian in the ordered phase $0.5237<J_2<1$,  scanning the two-dimensional
parameter space $\alpha$ --- $\beta_{\rm eff}$. Results are displayed in
Fig.~\ref{fig:min_div_alpha} where color maps of the normalized KL divergence $I_1/S_1$ are shown.
As in the gapped case (Fig.~\ref{fig:min_div_xi}), here we also see the development of the minimal
relative KL divergence, taking relatively small values $< 0.1\%$.  Interestingly, when $J_2$ decreases from 1, the minimum area moves very slowly, and only vertically, up to $J_2=0.6$. Indeed, in the ordered regime, we observe the optimal exponent $\alpha$ in the minimum region to remain almost constant $\sim 1.6$ (this will be discussed in more detail in Sec~\ref{sec:discussion}).  Interestingly it seems that the only varying parameter is the effective temperature which is found to slowly increase. Such a behavior is not surprising since one expects the quantum disordering of the N\'eel order in the dimerized model to translate into a thermal disordering for the entanglement Hamiltonian. 

It is also instructive to monitor the $\alpha$-model in the gapped regime, as shown in the bottom right panel of Fig.~\ref{fig:min_div_xi} for $J_2=0.5<J_c$. This panel is qualitatively distinct from the 5 others (in the N\'eel ordered regime) as the shape of the contour lines is different and the minimum position has moved horizontally towards $\alpha\sim 2$. This is a clear qualitative signature of the quantum phase transition between $J_2=0.6$ and $J_2=0.5$. Nevertheless, the $\alpha$-model still displays a very nice minimum in the disordered side, as we further discuss below in Sec.~\ref{sec:discussion}.

Let us come back now to the spectral comparison at $J_2=1$ displayed above in
Fig.~\ref{fig:spectra_lro}. Whereas the proximity of $\{\epsilon^E\}_{\alpha=1.65,\beta_{\rm eff}=1.47}$ and
$\{\epsilon^B\}_{J_2=1}$ appears quantitatively very satisfying, with a KL divergence
$I_1/S_1=1.218(1)\cdot10^{-3}$, we have to face  a couple of potential problems with the $\alpha$-model as an effective boundary entanglement Hamiltonian for the ordered phase. Firstly, a careful examination of the full PS on the left panel of  Fig.~\ref{fig:spectra_lro} leads to the observation that some local gaps in the middle of the spectrum $\{\epsilon^B\}$ are missing in the effective $\alpha$-model. Secondly, as observed in Refs.~\cite{Alba12,kolley_entanglement_2013}, an essential feature of the effective entanglement Hamiltonian is that it should also capture, on top of the TOS structure, the spin-wave excitations. A spin-wave analysis of the $\alpha$-model~\cite{yusuf_spin_2004,laflorencie_critical_2005} yields a long wave-length dispersion $\omega_{\rm sw}\sim k^{\frac{\alpha-1}{2}}$, and therefore a finite size gap $\Delta_{\rm sw}(L)\sim L^{\frac{1-\alpha}{2}}$. Nevertheless, we would expect the correct entanglement Hamiltonian ${\cal H}_E$ to display a low energy spectrum with SW excitation levels $\sim 1/L$, a requirement only possible if $\alpha=3$. Instead, the case $\alpha=1.65$ would lead to a wrong SW spectrum. Another argument comes from the fact that spin correlations fall as $1/r$ as a function of distance in the ground-state of the $\alpha-$ model Eq.~\ref{eq:alpha} (see Ref.~\cite{laflorencie_critical_2005}) when $\alpha=3$, the same dependence as in the 2d Heisenberg model on top of the long-range order.

In order to repair this inconsistency, we introduce another model (the so-called $\Lambda$-model), expected to be closer to the true entanglement Hamiltonian in the antiferromagnetic phase.

\subsubsection{$\Lambda$-model}

\begin{figure}[h]
     \centering
     \includegraphics[width=\columnwidth]{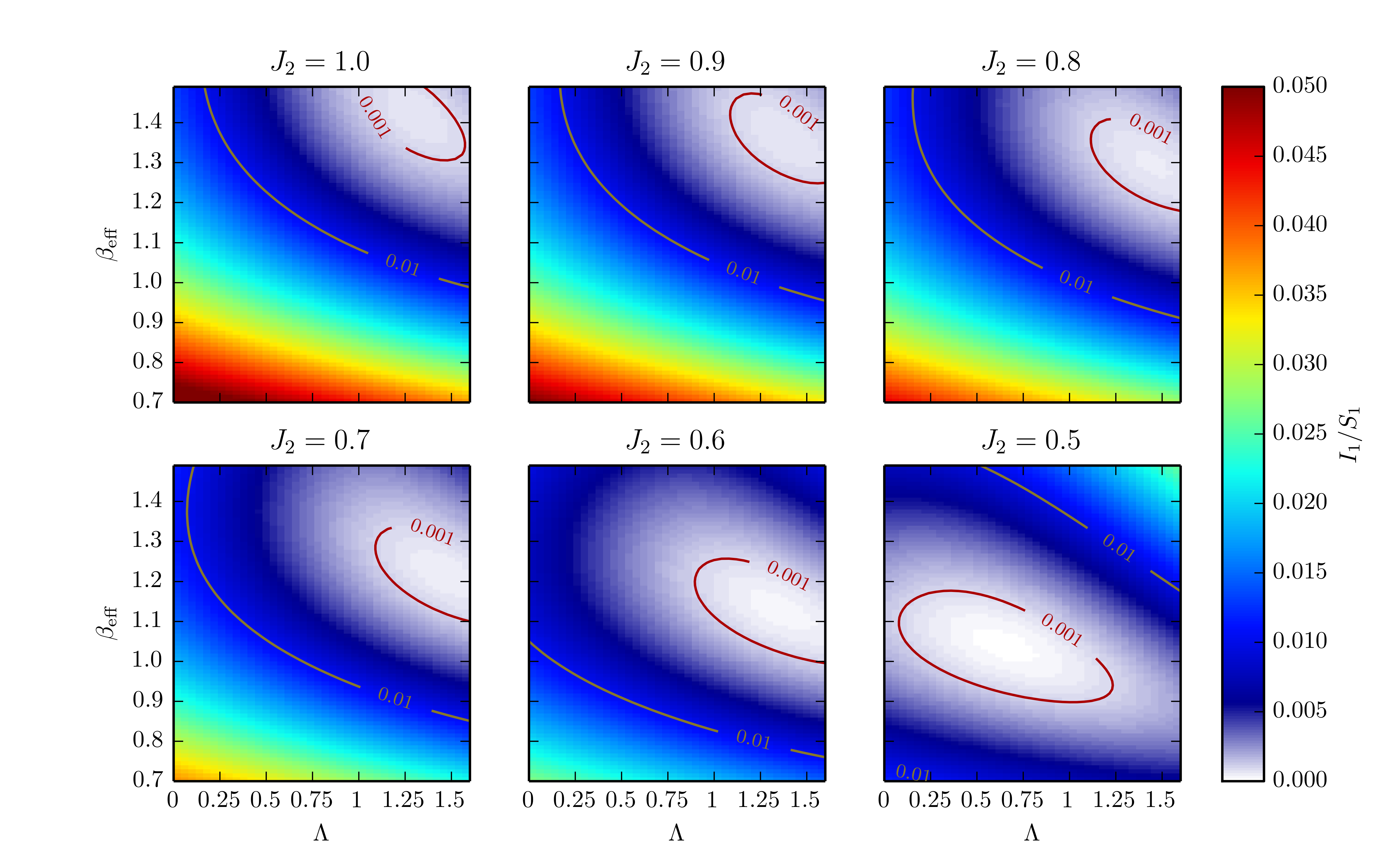}
     \caption{Relative KL divergences $I_1/S_1$ of the line shaped subsystem spectrum and the effective
     long range Hamiltonian given by the ``$\Lambda$-model'' (Eq. \ref{eq:alpha}) for different values
 of $J_2$ (here $L=16$). }
     \label{fig:div_B}
 \end{figure}

The Hamiltonian for the $\Lambda$-model is given by:
\be
{\cal H}_E(\Lambda)=\sum_{\stackrel{i,j\in B}{i>j}}{(-1)^{r_{ij}}} \left(\frac{\Lambda}{L}+\frac{1}{r_{ij}^{3}}\right)\vec{S}_i \cdot
\vec{S}_{j},
\label{eq:lambda}
\ee
where the $\Lambda$ term is constant for all $r_{ij}$, and the $1/L$ normalisation is necessary to
preserve the extensivity.  Such a Lieb-Mattis term does not sustain SW whereas the power-law
component $\sim 1/r^3$ is expected to bring $\omega_{\rm sw} \sim k$ SW excitations and $1/r$ decaying spin correlation functions. Again,
searching for the best  couple of parameters $(\Lambda,\beta_{\rm eff})$, we show in the left column
of the packets in  Fig.~\ref{fig:spectra_lro}~(a) that with $\Lambda=1.28$ and $\beta_{\rm
eff}=1.42$, the effective PS $\{\epsilon^E\}_{\Lambda,\beta_{\rm eff}}$ compares extremely well with
$\{\epsilon^B\}_{J_2=1}$, with a KL divergence as small as $I_1(\Lambda)/S_1 = 6.30(1) \cdot 10^{-4}$ and the small local gaps discussed above are now very well reproduced. Quantitatively speaking the $\Lambda$-model yields a better agreement, with smaller KL divergences for the optimal parameters, with $I_1/S_1 \sim 0.05\%$ across the entire N\'eel phase. In Fig.~\ref{fig:div_B} we see in the plane $\Lambda$ --- $\beta_{\rm eff}$  that the minimum keeps the same value of $\Lambda\sim 1.3-1.4$ while the effective temperature keeps increasing when N\'eel order is reduced.

\subsubsection{Discussion}
\label{sec:discussion}

 Finding the exact entanglement Hamiltonian is an exponentially complicated task which cannot be
 easily done in the general case. However, in many cases, guidance obtained by the knowledge of
 systems in dimension $d-1$,  symmetry considerations as well as physics that need to be described
 at finite temperature, one often has an intuition on which class of $d-1$ models could represent
 the entanglement Hamiltonian. In this case, the PS comparison approach, with a quantitative
 criteria provided by the KL or R\'enyi divergences, proves to be quite powerful.
 Note that the basis-dependence of the PS is no longer in issue as both the reduced density matrix and the (exponential of the) entanglement Hamiltonian are considered in the {\it same} basis: the exact entanglement Hamiltonian should indeed have the same PS (at finite effective temperature) than the subsystem considered, in any basis. We emphasize that the KL divergence Eq.~\ref{eq:I1} compares the two PS (of the subsystem and of ${\cal H}_E$) {\it basis state by basis state}: this is in contrast with the usual comparison of the ES low-lying levels where the eigenfunctions are discarded.

 Clearly, the fact that the PS of the subsystem and of an ansatz entanglement Hamiltonian are very
 close is not a full proof that the ansatz is indeed the exact entanglement Hamiltonian, as
 offdiagonal elements of the (reduced) density matrices could still be different. However, combined
 with physical arguments, it provides indirect evidence. Furthermore incorrect entanglement Hamiltonians can easily be ruled out if the PS do not match. 

\begin{figure}[h] 
     \centering
     \includegraphics[width=\columnwidth]{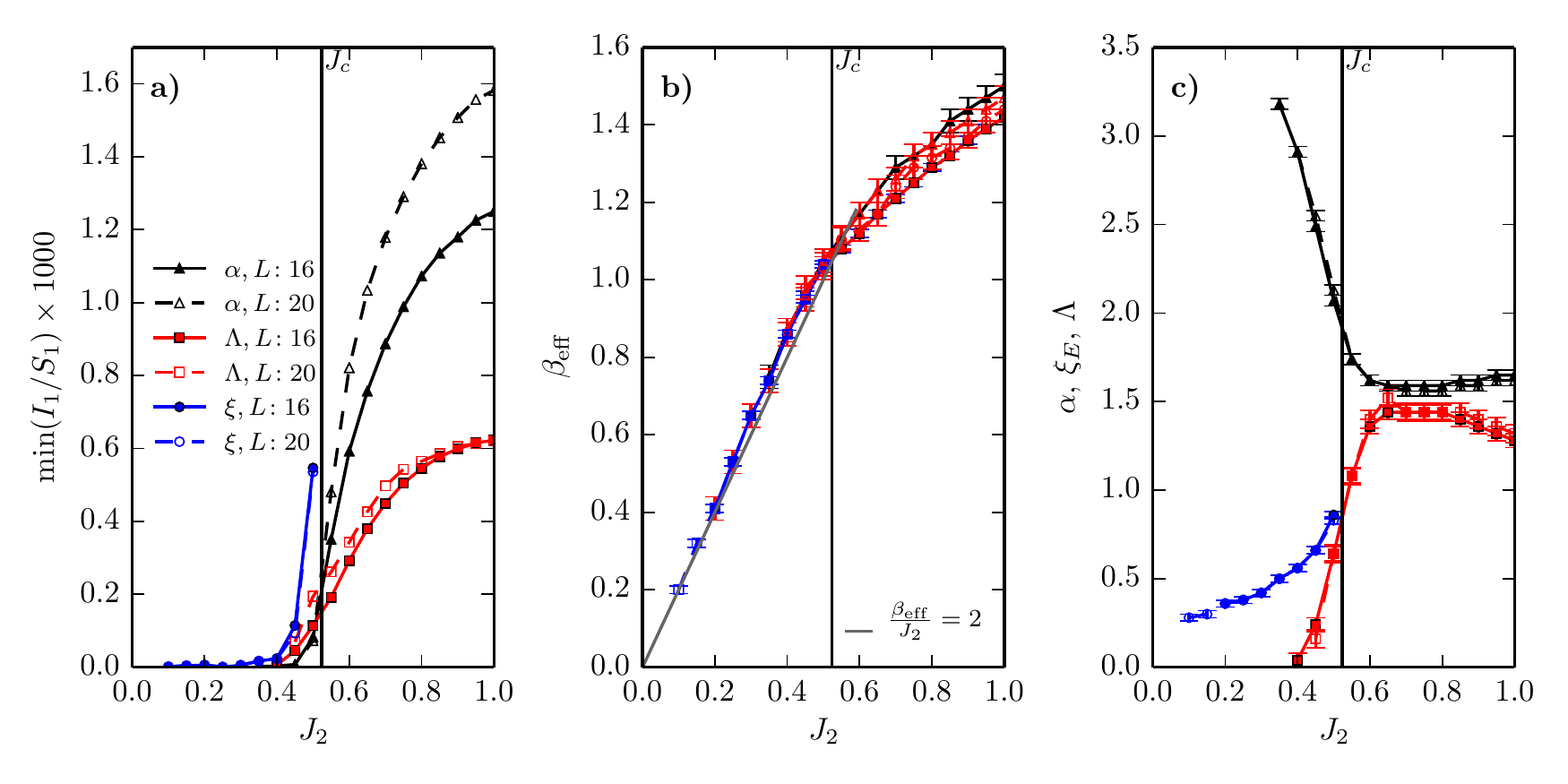}
     \caption{
(a) Comparison of the normalized KL divergence $I_1/S_1$ given in permils of the line shaped subsystem spectrum and
     the effective Hamiltonians for the optimal parameter set as a function of $J_2$. Errorbars from a bootstrap analysis for $I_1/S_1$ are typically smaller than $10^{-6}$, however due to the finite grid in parameter space, the uncertainty might be higher as the real minimum might lie between our sampled points.
(b) Effective inverse temperature of the different models. (c) Optimal parameters of the different models.
  }
     \label{fig:min_div_comp}
 \end{figure}

We summarize in Fig.~\ref{fig:min_div_comp} the results obtained using this approach for the
dimerized and N\'eel phases of the model Eq.~\ref{eq:Hdim}, as well as in the critical region
separating them, using the three ansatz entanglement Hamiltonians: Eq.~\ref{eq:xi} ($\xi$-model with
short-range Heisenberg interactions), Eq.~\ref{eq:alpha} ($\alpha-$model with power-law
interactions) and Eq.~\ref{eq:lambda} ($\Lambda-$model which contains a $1/r^3$ interaction as well
and a full-range interaction) for a line subsystem. In this figure, we display results obtained for a $L=16$ (corresponding to all the results presented so far) as well as for a $L=20$ subsystem, allowing to discuss size-dependence of our results. 

Deep in the dimerized phase, the $\xi$-model
with a very short range entanglement length-scale $\xi_E$ has a PS extremely close to the true PS, resulting in a very small KL divergence $I_1/S_1\sim 10^{-5}$ (Fig.~\ref{fig:min_div_comp} a). This is easily understood as the full system has indeed short-range correlations. When increasing $J_2$ up to $\sim 0.4$ the entanglement length-scale $\xi_E$ increases (Fig.~\ref{fig:min_div_comp} c), but much more slowly than the true correlation length (see Fig.~\ref{fig:xis}). The effective inverse temperature is linear with $J_2$, as correctly captured by the perturbative argument Eq.~\ref{eq:naive}. In this regime, the $\xi-$model results have almost no size dependence.

On the other side of the transition, PS in the N\'eel phase are best reproduced by the
$\Lambda$-model (Fig.~\ref{fig:min_div_comp} a), with an approximately constant
(Fig.~\ref{fig:min_div_comp} c) full-range $\Lambda$ term and a finite effective temperature, slowly
decreasing with increasing N\'eel order. The power-law model with an exponent $\alpha \sim 1.6$
(Fig.~\ref{fig:min_div_comp} c) gives also a less precise, albeit reasonable, modeling of the PS (at
the hand-waving level, one might consider that $\alpha \sim 1.6$ is the best compromise for a pure
power-law interaction for mimicking the sum of the $\alpha=0$ and $\alpha=3$ terms in the
$\Lambda$-model). The long-range nature of the interactions of these models are also consistent with
the fact that the full system orders. The effective temperature decreases with $J_2$
(Fig.~\ref{fig:min_div_comp} b), consistent with an increased staggered magnetization on the bulk
system. The difference between the $L=16$ and $L=20$ data also confirms that the
$\Lambda-$ model describes more faithfully the true participation spectra: the $L=20$ data shows a
clearly larger relative KL divergence $I_1/S_1$ than the $L=16$ data for the $\alpha-$ model,
whereas there is only a very small loss of precision when going from $L=16$ to $L=20$ with the
$\Lambda-$ model.
  It should be noted that the statistical uncertainty of our results for the relative KL divergences
is extremely small due to very high precision of the histograms in combination with a large
precision gain by the exploitation of spatial symmetries. However, due to the finite grid in
parameter space used for the estimation of the point with minimal KL divergence, a systematic error
remains which might partly explain the slight difference of the $L=20$ and $L=16$ divergences for
the $\Lambda$ model.

Close to the quantum phase transition in the disordered phase (in the range $0.4-J_c$ for our
finite-size simulations), it appears that the $\alpha-$ and $\Lambda-$ models provide a smaller KL
divergence than the $\xi-$ model. We would like to emphasize that this is {\it not} in contradiction
with the fact that the bulk system is disordered: indeed in this region, the optimal power-law
exponent $\alpha \in [2.0,2.5]$, the decreased full range interaction $\Lambda$
(Fig.~\ref{fig:min_div_comp} c) and the increased optimal effective temperature
(Fig.~\ref{fig:min_div_comp} b) are such that the $\alpha-$ and $\Lambda-$ models are in their {\it
disordered phase}. It is well possible that increasing the subsystem size will result in inversion of
the KL divergences curves between the $\alpha-$, $\Lambda-$ and the $\xi-$ models (indeed, on a
finite system of length comparable with the bulk correlation length, the system appears ordered) --
this possibility is already compatible with the finite-size trend observed between the $L=16$ and
$L=20$ data. We finally remark that at or close to the quantum critical point, the $\alpha-$ model
with $\alpha \sim 2$ appears more optimal than the $\Lambda-$ model, indicating that the critical
entanglement Hamiltonian may have pure power-law interactions (even though the nature of ${\cal
H}_E$ is difficult to capture due to increased finite-size effects close or at the quantum critical
point).

We would like to emphasize that the true ${\cal H}_E$ certainly contains more complex, multi-spin interactions, which are not included in our present analysis with the ansatz $\xi-$, $\alpha-$ and $\Lambda-$ models, especially in the ordered and critical parts of the phase diagram. Nonetheless, we find that the very small KL divergences observed (less than one per thousand) for the optimal models already indicate that the leading two-spin part of ${\cal H}_E$ is reasonably well captured by these ansatz models.  

Finally, it is important to note that the 1d bipartition studied in this paper necessarily implies a
non-vanishing effective temperature $T_{\rm eff}$, even in the thermodynamic limit. Indeed, for such
line shaped subsystems, area and volume have similar scalings. The thermal entropy of the effective
model $S_{E}^{\rm therm}(L,T_{\rm eff})$ being equal to the zero temperature von-Neumann EE of subsystem B $S_{B}^{\rm vN}(L)$,  they both scale with $L$ and therefore, a vanishing $T_{\rm eff}(L)$ with $L$ would imply a wrong sublinear scaling of $S_{E}^{\rm therm}$ and $S_{B}^{\rm vN}$.

\newpage
\section{Conclusions}

The analysis presented in the last part of this paper (Sec.~\ref{sec:sub}), shows the usefulness of
studying participation spectra for understanding which entanglement Hamiltonian can emerge at the
boundary of a physical system cut by a bipartition. While this approach cannot provide {\it per se}
the entanglement Hamiltonian, it allows to rapidly test whether a physically-motivated ansatz
entanglement model correctly describes the subsystem physics at play.  An interesting aspect is that
effects of the entanglement temperature can directly be considered in this approach, which is
clearly different from an analysis based on solely the (low-lying) levels of the entanglement
spectrum. In addition, we also introduced in Sec.~\ref{sec:KL} the Kullback-Leibler and R\'enyi
divergences which provide a quantitative way (a number) to characterize how close a physical
(entanglement or participation) spectrum is from the one of a reference system. This will certainly
be useful for methods which provide a direct access to entanglement spectra. Note that a visual
inspection of spectra is easily misguided as different states may have different multiplicities and
the representation in the form of spectra usually does not distinguish the corresponding states.
Both problems are easily solved by the introduction of KL and R\'enyi divergences.
Finally, simple arguments (Sec.~\ref{sec:gap}) based on the existence of a gap in the PS allow to provide an exact proof for the existence of a gap in the ES of Rokshar-Kivelson wave-functions.

There are still many open questions regarding the physical information contained in SR entropies and
participation spectra, some of which have been discussed in the review part Sec.~\ref{sec:review} of
this paper, highlighting a clear need of analytical, field-theoretical work, to understand some of the finite-size scalings observed in numerical simulations of the SR entropies. Also, the connection with similar quantities in classical models at criticality will be useful to discuss universal behavior. The existence of universal subleading terms, as well as the behavior of the leading term (as a function of the Hamiltonian parameters) deserve to be studied for other phases of matter, such as topological phases for instance. Another appealing aspect is to understand the behavior of SR entropies at {\it finite} temperature, {\it i.e.} beyond the ground-
state studies discussed in this work. In all these cases, the possibility to use on fairly large systems, a stochastic Monte Carlo approach~\cite{luitz_universal_2014}, will certainly turn crucial.

\section{Acknowledgments}
We would like to foremost thank J.-M. St\'ephan and G. Misguich for many suggestions which improved this manuscript as well as for illuminating discussions, in particular for explaining the predictions of boundary CFT. We also thank G. Misguich, M. Oshikawa, X. Plat and I. McCulloch for collaborations on this topic. Useful discussions with C. Sire, B. Georgeot, G. Lemari\'e and G. Roux are also acknowledged. QMC simulations were performed codes using numerical resources from GENCI (grants 2013-x2013050225 and 2014-x2014050225) and CALMIP. Our codes are partly based on the ALPS libraries~\cite{ALPS13,ALPS2}. This work is supported by the French ANR program ANR-11-IS04-005-01.

\newpage
\bibliographystyle{unsrt}
\def\url#1{} \def\urlprefix{}

\end{document}